\documentclass[aps,prd,twocolumn,groupedaddress,nofootinbib,floatfix,]{revtex4-2}
\usepackage{booktabs}\usepackage{xcolor}
\usepackage{amsmath}
\usepackage{array}
\usepackage{amssymb}
\usepackage{graphicx}
\usepackage{hyperref}
\usepackage{mathtools}
\usepackage{orcidlink}
\usepackage{natbib}
\usepackage{hyperref}
\usepackage{multirow}
\usepackage{bm}

\def \be {\begin{equation}}
\def \ee {\end{equation}}
\def \ba {\begin{array}}
\def \ea {\end{array}}
\def \bea{\begin{eqnarray}}
\def \eea{\end{eqnarray}}

\newcommand{\NCUa}{Department of physics, Nanchang University, Nanchang, 330031, China}
\newcommand{\NCUb}{Center for Relativistic Astrophysics and High Energy Physics, Nanchang University, Nanchang, 330031, China}

\newcommand{\dd}{\mathrm{d}}
\newcommand{\avg}[1]{\left\langle #1\right\rangle}

\newcommand{\beq}{\begin{equation}}
\newcommand{\eeq}{\end{equation}}

\providecommand{\ac}[1]{#1}
\providecommand{\ac}[1]{#1}
\begin{document}
\title{Beyond circular and equatorial orbits: Bayesian constraint on vector charge with the adiabatic inspirals of generic Kerr geodesics}

\author{Tieguang Zi\,\orcidlink{0000-0003-0046-2056}}
\affiliation{\NCUa}
\affiliation{\NCUb}

\author{Genliang Li\,\orcidlink{0009-0008-9893-6160}}
\email{ligenliang20@mails.ucas.ac.cn}
\affiliation{College of Computer and Big data, Putian University, Putian 351100, Fujian, China}

\begin{abstract}
Extreme-mass-ratio inspirals (EMRIs) offer a powerful probe of additional fundamental fields through their cumulative influence on gravitational-wave phases. However, modeling vector radiation
from generic Kerr orbits requires the evolution of the Carter constant, which cannot be determined from energy and angular-momentum balance alone. Here we derive the orbit-averaged electromagnetic
Carter-constant flux for eccentric and inclined, nonresonant Kerr geodesics, including sections at infinity and event horizon. Combining the gravitational and electromagnetic energy and angular-momentum fluxes, we construct generic adiabatic inspirals using a four-dimensional Chebyshev interpolation method. We incorporate the relativistic trajectories into the state of the art of \texttt{FastEMRIWaveform} package and perform Bayesian inference of vector charge carried by secondaries. Our analysis reveals substantial correlations between the charge and intrinsic source parameters, with vector charge constraints exhibiting a nonmonotonic dependence on orbital inclination. These results establish a framework for investigating additional vector-radiation channels in generic EMRIs and their implications for tests of gravity with future space-based gravitational-wave observations.
\end{abstract}

\maketitle

\section{Introduction}
Compact-binary coalescences have become the accurate laboratories for testing relativistic gravities, providing access to dynamical phenomena that cannot be reconstructed from electromagnetic observations alone~\cite{Bailes:2021tot,Meszaros:2019xej}. The LIGO--Virgo--KAGRA (LVK) network currently probes the  gravitational-wave (GW) signal from the frequency band of several tens to several thousand hertz, where signals from the coalescences of stellar-mass compact binaries carry information about the source masses, spins and orbital dynamics throughout the late inspiral, merger and ringdown~\cite{KAGRA:2021vkt}.  In particular, the post-merger signal provides a direct probe of the spacetime of the remnant black hole. Its characteristic nature, together with the nonmodal features in GW data stream,  allows us to test the deviation of black-hole spectroscopy and the Kerr paradigm in the strong-field region~\cite{Kokkotas:1999bd,Dreyer:2003bv,Berti:2009kk,Cardoso:2019rvt,Hughes:2019zmt,Berti:2025hly,Jaramillo:2020tuu,Jaramillo:2022kuv,Destounis:2021lum,Zi:2021pdp,Destounis:2023nmb,Sarkar:2023rhp,Boyanov:2022ark,Boyanov:2023qqf,Cheung:2021bol,Chen:2026ehv,Boyanov:2024fgc,Cai:2025irl,Rosato:2024arw,Ianniccari:2024ysv,dePaula:2025fqt,Wang:2025rvn,Xiong:2024urw,Rosato:2025lxb}. As detector sensitivities and waveform models continue to improve, GW observations are becoming increasingly powerful probes of both compact-object astrophysics and possible departures from general relativity (GR).

A complementary view of testing relativistic dynamics will be provided by the GW signal in the millihertz GW band, where massive compact binaries evolve sufficiently slowly, their inspirals can last months or even years.  This frequency range will be explored by space-based interferometers such as LISA~\cite{LISA:2017pwj}, TianQin~\cite{TianQin:2020hid,Li:2024rnk} and Taiji~\cite{Ruan:2018tsw,Du:2025xdq}, whose target source populations include galactic compact binaries, massive-black-hole binaries and extreme-mass-ratio inspirals (EMRIs)~\cite{Barausse:2020rsu,LISA:2022yao,LISA:2022kgy,Karnesis:2022vdp}. EMRIs are especially compelling because their extreme mass-ratio, $\epsilon=10^{-4}$--$10^{-7}$, the secondary can consequently execute $\sim10^4$--$10^5$ orbital cycles in the strong-field region of the massive black hole (MBH) before the final plunge.  The accumulated GW signal acts as a coherent record of the orbital dynamics: even perturbatively small corrections to the spacetime background can generate the measurable modifications. A typical EMRI has the long-duration inspiral phase in the strong-field region, allowing to probe the Kerr nature of the central object and additional interactions associated with new fields or the astrophysical environment~\cite{Amaro-Seoane:2007osp,Babak:2017tow,Berry:2019wgg,Cardenas-Avendano:2024mqp,Barack:2003fp,Baibhav:2019rsa}.

Extreme- and intermediate-mass-ratio binaries provide a particularly sensitive arena for probing gravity in the strong-field regime~\cite{Cardenas-Avendano:2024mqp,Barack:2018yly,Barausse:2020rsu,
Barausse:2016eii,Blazquez-Salcedo:2016enn,Glampedakis:2005cf,Barack:2006pq,Cardoso:2018zhm,Datta:2019epe,Pani:2019cyc,Maggio:2021uge,Destounis:2020kss,Piovano:2020ooe,Sago:2021iku,Piovano:2022ojl,Zi:2023geb,Zi:2023pvl,Zi:2021pdp,Zi:2022hcc,Zi:2024itp,Zi:2024lmt,Zi:2025jxy,Zi:2026zpw,Zi:2026cwm,Li:2026hyn,Zi:2025idv,Du:2026xre}. Exploiting this potential of EMRI, however, requires waveform models capable of tracking the orbital phase with high accuracy over the large number of cycles accumulated during the inspiral. Within general relativity (GR), the small mass-ratio $\epsilon=m_p/M\ll1$ provides a natural perturbative parameter, and the gravitational self-force (SF) program offers a systematic framework for describing the motion of the secondary and the associated gravitational radiation~\cite{Barack:2018yvs,Pound:2019lzj,Warburton:2021kwk,Wardell:2021fyy}.
The corresponding problem is considerably more involved in theories beyond GR, where additional propagating degrees of freedom may modify both the conservative dynamics and the dissipative sector.  Although rotating black-hole solutions have been constructed in a number of alternative theories~\cite{Yunes:2009hc,Pani:2011gy,Yagi:2012ya,Kleihaus:2015aje,Cunha:2019dwb,Delgado:2020rev}, their specific forms are generally theory-dependent, it would be very difficult to construct a suitable and universal perturbative theory describing the relativistic two-body problem.  Considerable progress has nevertheless been made toward extending the Kerr perturbation framework beyond GR, including formulations in which Teukolsky-like equations are generalized to classes of modified theories of gravity~\cite{Wagle:2023fwl,Li:2022pcy}.

A particularly useful route for treating small-mass-ratio binaries beyond GR is provided by effective-field-theory and skeletonized descriptions of EMRIs. In the framework developed in Ref.~\cite{Maselli:2020zgv}, EMRI as the probe of detecting fundamental fields beyond the \ac{GR} have been broadly investigated in the Refs. \cite{Maselli:2020zgv,Maselli:2021men,Barsanti:2022ana,Barsanti:2022vvl,Zhang:2022rfr,Zhang:2023vok,
Fell:2023mtf,DellaRocca:2024pnm,Kumar:2024utz,Zi:2026cwm,Barsanti:2026ulr,Speri:2024qak,Gliorio:2026yvh}, where the MBH can be described by the Kerr metric according to the no-hair theorem and the stellar-mass \ac{BH}s can carry the scalar charges that are dependent on their masses~\cite{Mignemi:1992nt,Kleihaus:2011tg,Antoniou:2017acq,Antoniou:2017hxj,Doneva:2017bvd,Maselli:2020zgv}. Such an assumption is based on the conclusion that the charge of BH is weaker when it becomes more massive. It is due to that the charges are controlled by the curvature, and when the mass of BH is increasing the curvature near the horizon tends to diminish \cite{Maselli:2020zgv}. Therefore, the MBH would behave the much weaker deviation of Kerr hypothesis, and the stellar-mass BHs may occur the violation of the no-hair theorem. The dominant beyond-GR effect is then encoded in scalar charge of the secondary, which sources an additional radiative channel and modifies the secular evolution of the orbit.  This construction is especially attractive from the viewpoint of waveform modeling because it preserves much of the standard Kerr perturbation machinery while introducing only a small number of additional parameters.  In its simplest realization, the leading modification is controlled by the scalar charge or vector charge of the secondary. The same physical setup has subsequently been incorporated into a self-force expansion, allowing post-adiabatic corrections to the orbital dynamics to be treated systematically~\cite{Spiers:2023cva}.

The applicability of this framework has been progressively extended beyond the simplest configurations of circular and equatorial orbits.  The orbital evolution derived by scalar radiation have been investigated for equatorial eccentric orbits~\cite{Barsanti:2022ana,Zhang:2022rfr}, spherical inclined orbits~\cite{DellaRocca:2024pnm}, and, recently, fully generic bound Kerr orbits~\cite{DellaRocca:2024pnm,Zi:2025lio,Gliorio:2026yvh}.  Further developments have considered massive
scalar fields in circular configurations~\cite{Barsanti:2022vvl} and
time-dependent scalar charges~\cite{DellaRocca:2024sda}.  Complementary
scalar self-force calculations have also been performed for equatorial
circular~\cite{Warburton:2010eq}, equatorial eccentric
~\cite{Warburton:2011hp}, spherical inclined
~\cite{Warburton:2014bya}, and generic bound trajectories
~\cite{Nasipak:2019hxh}.  These studies have established a progressively
more complete description of scalar-field effects across the orbital
parameter space relevant to EMRIs.  At the waveform level, several analyses
have further demonstrated that even perturbatively small scalar charges can
produce cumulative phase corrections that become potentially observable
over the long duration of an EMRI signal
~\cite{Maselli:2021men,Speri:2024qak}.

More recently, the same perturbative strategy has begun to be generalized from the scalar to vector fields.  Vector radiation from a vector-charged secondary has been studied for circular and eccentric trajectories~\cite{Zhang:2023vok,Torres:2020fye,German:2023bye,Zi:2025jxy,Zi:2025qos}, providing the first steps toward incorporating additional spin-$1$ dissipative radiation into EMRI dynamics.  Despite this progress, the vector sector remains considerably less developed for generic Kerr geodesics.  In particular, a generic bound orbit with three independent radial, polar and azimuthal orbital frequencies possesses the rich relativistic effects in the strong gravitational field, and its secular evolution requires the simultaneous change rate of the energy, angular-momentum and Carter-constant. In this paper, using these rates of change  of orbital integrals of motion, we take the first step to evolve the generic geodesic orbits influenced by both gravitational and electromagnetic radiation, then present the rigorous constraint on vector charge using Bayesian analysis.

The structure of this work is arranged as follows: we introduce the method of the change rate of Carter constant and constraint on vector charge using EMRI waveform from the inclined and eccentric orbits in Sec.~\ref{method}, we first show the theoretical framework of describing EMRI system in the family of Einstein-Maxwell theory in Sec.~\ref{sec:setup0} and the geodesic equation of Kerr spacetime in Sec.~\ref{sec:setup}. The Sec.~\ref{sec:scalarization} shows the derivation process of Carter flux driven by vector perturbation in Kerr spacetime and we introduce the adiabatic evolution method of orbital parameters in Sec.~\ref{sec:evolution}.
We present our results of EMRI energy fluxes on the generic orbits and the MCMC constraint on vector charge using different orbital parameter configurations in Sec.~\ref{sec:result}. Finally, we  summarize the main results in this work and give some discussion in Sec.~\ref{sec:conclusion}. Throughout this paper, we use geometrized units $G=c=4\pi\epsilon_0=1$ and metric signature $(-,+,+,+)$.  

\section{Method}\label{method}
\subsection{Setup}
\label{sec:setup0}

We consider a broad class of theories described by the action
\begin{equation}
S[g,A,\Psi]
=
S_0[g,A]
+
\alpha S_c[g,A,\Psi]
+
S_m[g,A,\Psi],
\label{action}
\end{equation}
where $g_{\mu\nu}$ is the spacetime metric, $A_\mu$ denotes a massless vector field, and $\Psi$ collectively represents the remaining matter degrees of freedom.  The term $S_0[g,A]$ describes the Einstein--Maxwell sector, whereas $S_c[g,A,\Psi]$ contains possible nonminimal interactions between the gravitational, vector and matter
fields.  The coupling constant $\alpha$ has mass dimension $[\alpha]=({\rm mass})^n, ~ n>0,$ while $S_m[g,A,\Psi]$ denotes the matter action. The Einstein--Maxwell sector is taken as
\begin{equation}
S_0[g,A]
=
\int d^4x\,\sqrt{-g}
\left[
\frac{R}{16\pi}
-
\frac{1}{4}F_{\mu\nu}F^{\mu\nu}
\right],
\label{Lagrangian}
\end{equation}
where $R$ is the Ricci scalar associated with $g_{\mu\nu}$ and
\begin{equation}
F_{\mu\nu}
=
\nabla_\mu A_\nu-\nabla_\nu A_\mu
\end{equation}
is the Maxwell field-strength tensor.  The coupling of the vector field
to the compact secondary is included in the matter sector and gives rise
to the current $J^\mu$.

Variation of the total action with respect to $g_{\mu\nu}$ and $A_\mu$
yields, schematically,
\begin{align}
G_{\mu\nu}
&=8\pi\left(T_{\mu\nu}^{\rm vec}
+T_{\mu\nu}^{p}+\alpha T_{\mu\nu}^{c}\right),
\label{eq:metric}
\\
\nabla_\rho F^{\rho\mu}
+
\frac{4\pi\alpha}{\sqrt{-g}}
\frac{\delta S_c}{\delta A_\mu}
&=
4\pi J^\mu ,
\label{eq:vector_full}
\end{align}
where $T_{\mu\nu}^{\rm vec}$ is the stress-energy tensor of the vector
field, $T_{\mu\nu}^{p}$ denotes the stress-energy tensor of the compact
secondary, and $T_{\mu\nu}^{c}$ contains the contribution associated with
the nonminimal interaction $S_c$. For a pointlike secondary of mass $m_p$ carrying dimensionless vector
charge $q_v$, the corresponding current is
\begin{equation}
J^\mu(x)
=
q_v m_p
\int d\tau\,
u^\mu
\frac{
\delta^{(4)}[x-y(\tau)]
}{
\sqrt{-g}
},
\label{eq:vector_current}
\end{equation}
where $y^\mu(\tau)$ denotes the particle worldline,
$u^\mu=dy^\mu/d\tau$ is its four-velocity, and $\tau$ is the proper time.
The symbol $\delta^{(4)}[\cdot]$ denotes the four-dimensional Dirac delta
distribution.

We further simplify the field equations in the framework of EMRI, then assume that the primary black hole carries no a negligible vector charge. We next introduce the perturbative expansion appropriate to a typical EMRI system. Defining
\begin{equation}
\epsilon\equiv\frac{m_p}{M}\ll1,
\end{equation}
where $M$ is the mass of the primary black hole, the dimensionless
coupling controlling deviations from general relativity can be written as
\begin{equation}
\xi
\equiv
\frac{\alpha}{M^n}
=
\epsilon^n
\frac{\alpha}{m_p^n}.
\label{eq:xi_def}
\end{equation}
We adopt the small-coupling quantity $\frac{\alpha}{m_p^n}\lesssim
{\cal O}(1),$ so that $\xi\lesssim{\cal O}(\epsilon^n)$.  This assumption provides a
controlled perturbative regime in which modifications of the background
geometry are parametrically suppressed in the extreme-mass-ratio limit.
It is also consistent with the absence of compelling observational
evidence for large deviations from the Kerr geometry in either weak- or strong-field tests of gravity~\cite{Will:2014kxa,LIGOScientific:2025cmm,LIGOScientific:2025rid}.

At zeroth order in the mass ratio, the spacetime is
therefore described by the Kerr metric and the background vector field
may be chosen to vanish,
\begin{equation}
g_{\mu\nu}^{(0)}=g_{\mu\nu}^{\rm Kerr},
\qquad
A_\mu^{(0)}=\rm constant .
\label{eq:background_fields}
\end{equation}
The metric and vector fields can then be expanded in powers of the mass
ratio as
\begin{align}
g_{\mu\nu}
&=
g_{\mu\nu}^{(0)}
+
\epsilon h_{\mu\nu}^{(1)}
+
{\cal O}(\epsilon^2),
\nonumber\\
A_\mu
&=
\epsilon A_\mu^{(1)}
+
{\cal O}(\epsilon^2).
\label{eq:field_expansions}
\end{align}

At first order in $\epsilon$, the gravitational and vector perturbations
are sourced by the compact secondary.  In the skeletonized description
~\cite{Ramazanoglu:2016kul,Maselli:2020zgv,Maselli:2021men,Barsanti:2022vvl}, the extended matter sector is replaced
by an effective point-particle action.  For the minimally coupled
massless-vector sector considered here, it may be written as
\begin{equation}
S_p
=
-m_p\int d\tau
+
q_v m_p
\int A_\mu\,dy^\mu .
\label{eq:particle_action}
\end{equation}
The second term generates the current in
Eq.~\eqref{eq:vector_current}.  More general skeletonized descriptions
may additionally allow the effective mass to depend on local scalar
invariants constructed from the vector field.

In a locally inertial frame
$\tilde{x}^{\mu}$ centered on the secondary, the leading near-zone
behavior of the vector perturbation has the Coulomb form
\begin{equation}
A_{\tilde t}^{(1)}
=
\frac{q_v m_p}{\tilde r}
+
{\cal O}
\left(
\frac{m_p^2}{\tilde r^2}
\right),
\label{eq:local_vector}
\end{equation}
up to convention-dependent signs and gauge transformations, while the
spatial components are subleading in the instantaneous rest frame.

An important consequence of the expansion~\eqref{eq:field_expansions} is that the vector stress-energy tensor is quadratic in the first-order vector perturbation $T_{\mu\nu}^{\rm vec}={\cal O}(\epsilon^2)$.
Moreover, provided that the nonminimal sector vanishes on the Kerr
background and that its leading variation with respect to the vector
field is linear in $A_\mu^{(1)}$, its contribution to the vector equation
scales as
\begin{equation}
\alpha
\frac{\delta S_c}{\delta A_\mu}
=
{\cal O}(\epsilon^{n+1}),
\end{equation}
and is therefore subleading with respect to the
${\cal O}(\epsilon)$ point-particle source for $n\geq1$.  Under these
assumptions, the leading-order perturbation equations reduce to
\begin{align}
G_{\mu\nu}^{(1)}[h]
&=
8\pi m_p
\int d\tau\,
u_\mu u_\nu
\frac{
\delta^{(4)}[x-y(\tau)]
}{
\sqrt{-g^{(0)}}
},
\label{eq:einstein}
\\
\nabla^{(0)\rho}
F_{\rho\mu}^{(1)}
&=
4\pi q_v m_p
\int d\tau\,
u_\mu
\frac{
\delta^{(4)}[x-y(\tau)]
}{
\sqrt{-g^{(0)}}
}.
\label{eq:vector}
\end{align}
All covariant derivatives in Eqs.~\eqref{eq:einstein} and
\eqref{eq:vector} are taken with respect to the Kerr background.
The quantity $G_{\mu\nu}^{(1)}[h]$ denotes the linearized Einstein tensor
constructed from $h_{\mu\nu}^{(1)}$, whereas
$F_{\mu\nu}^{(1)}$ is the field strength associated with
$A_\mu^{(1)}$.

Both perturbation equations can be treated within the Kerr black-hole
perturbation framework~\cite{Teukolsky:1973ha}.  The gravitational sector
is conveniently described by the spin-$2$ Teukolsky equation, while the
vector sector is governed by the corresponding spin-$1$ Teukolsky
equation.  The gravitational fluxes for generic Kerr geodesics can be
computed efficiently using existing implementations in the
\texttt{Black Hole Perturbation Toolkit} (\texttt{BHPT})~\cite{BHPToolkit}.  In the
following sections, we focus on the spin-$1$ sector and compute the
electromagnetic energy, angular-momentum and Carter fluxes generated by a
vector-charged secondary.  These fluxes, together with their
gravitational counterparts, provide the dissipative contribution required to
evolve generic EMRI orbits in the adiabatic approximation.

\subsection{Kerr Spacetime and geodesic orbits}\label{sec:setup}
In the Boyer-Lindquist coordinates $\{t,r,\theta,\phi\}$, the Kerr solution is written as 
\begin{equation}
\begin{aligned}
ds^2 &= g_{\mu\nu} dx^\mu dx^\nu   \\
&=-\frac{ \Delta}{\Sigma}(dt - a\sin^2\theta d\phi)^2 + \frac{\Sigma}{\Delta}dr^2+\Sigma d\theta^2 \\
&+ \frac{\sin^2\theta}{\Sigma} \Big[(r^2+a^2)d\phi-a dt \Big]^2,
\end{aligned}
\end{equation}
with
\begin{equation}
 \Delta=r^2-2Mr+a^2,
 \qquad
 \Sigma=r^2+a^2\cos^2\theta,
 \label{eq:DeltaSigma}
\end{equation}
the event horizon is located at $r_+=M+\sqrt{M^2-a^2}$ and its angular velocity is  $\Omega_H=a/2Mr_+$. Here $a$ is the spin parameter of Kerr MBH, which has a dimension of MBH-mass $M$. 

Assuming that $x^\mu_p(\tau)$ is the particle's worldline, the tangent vector $u^\mu=\frac{dx_p^\mu}{d\tau}$ should have the following normalization  relationship $g_{\mu\nu}u^\mu u^\nu=-1$, four-velocity can be given by
\begin{equation}
 (u_t, u_r, u_\theta, u_\phi)
 =\left(-E,\,\sigma_r\frac{\sqrt{R(r)}}{\Delta},\,
 \sigma_\theta\sqrt{\Theta(\theta)},\,L_z\right),
 \label{eq:uext}
\end{equation}
where $\sigma_r,\sigma_\theta=\pm1$ are chosen on the inclined and eccentric geodesic orbit's turning points for radial and polar motion, functions $R(r)$ and $\Theta(\theta)$ are determined by  Eqs.\ \eqref{eq:geodesic:R}--\eqref{eq:geodesic:phi}. Geodesic orbits in the Kerr spacetime, satisfying $u^\nu \nabla_\nu u^\mu=0$, are characterized by three constants of motion, which are energy $E$ and angular momentum $L_z$,  normalized with the mass $m_p$  of secondary body, and Carter constant $Q=K-(L_z-a E)^2$, normalized with $m_p^2$, defined as
\begin{equation}
E=-u_\mu\xi^\mu_{(t)},~ L_z=u_\mu\xi^\mu_{(\phi)},
~K = K^{\mu\nu} u_\mu u_\nu.
\end{equation}
Here $\xi^\mu_{(t)} = (\partial_t)^\mu$ and $\xi^\mu_{(\phi)} = (\partial_\phi)^\mu$ are Killing vectors, $K$ is the variant of Carter constant and $K^{\mu\nu}$ is the Killing tensor, which is defined by 
\begin{equation}\label{eq:KillingTensor}
K_{\mu\nu}=2\Sigma l_{(\mu}n_{_\nu)} + r^2 g_{\mu\nu}
\end{equation}
where $l_\mu$ and $n_{\nu}$ are principal null vectors of the Kinnersley tetrad.
Four null basis $\{l^\mu,n^\mu,m^\mu,\bar{m}^\mu\}$ can determine the inverse of Kerr metric, 
\begin{equation}
g^{\mu\nu}=-2l^{(\mu}n^{\nu)} +2 m^{(\mu}\bar{m}^{\nu)}\;,
\end{equation}
where the Kinnersley tetrad are
\begin{align}
l^\mu_\pm &\equiv \left[ \pm (r^2+a^2) / \Delta, 1, 0, \pm a / \Delta \right] , \\
m^\mu_{\pm} &\equiv \left[ \pm i a \sin \theta, 0, 1, \pm i \csc \theta \right] = \overline{m}^\mu_{\mp}\;,\\
n^\mu &=-\frac{\Delta}{2\Sigma}l^\mu_-\;, m^\mu =-\frac{1}{\sqrt{2}(r+ia\cos\theta)}m^\mu_+\;.
\label{eq:tetrad2}
\end{align}

Using the constants of motion and introducing Mino time $\lambda$ defined by $\dd\tau=\Sigma\,\dd\lambda$~\cite{Mino:2003yg}, 
the geodesic equations are written as
\begin{align}\label{eq:geodesic:R}
 \left(\frac{\dd r}{\dd\lambda}\right)^2&=R(r),\\
 \left(\frac{\dd\theta}{\dd\lambda}\right)^2&=\Theta(\theta),  \label{eq:geodesic:Theta}\\
 \frac{\dd t}{\dd\lambda}&=T_r(r)+T_\theta(\theta), \label{eq:geodesic:t}\\
 \frac{\dd\phi}{\dd\lambda}&=\Phi_r(r)+\Phi_\theta(\theta), \label{eq:geodesic:phi}
\end{align}
with
\begin{align}
 P(r)&=E(r^2+a^2)-aL_z,\\
 R(r)&=P(r)^2-\Delta\left(r^2+K\right),
 \label{eq:R}\end{align}
and
\begin{equation}
 \Theta(\theta)
 =Q-\cos^2\theta
 \left[a^2(1-E^2)+L_z^2\csc^2\theta\right].
 \label{eq:Theta}
\end{equation}
\begin{align}
 T_r&=\frac{(r^2+a^2)P}{\Delta},
 &T_\theta&=a\left(L_z-aE\sin^2\theta\right),\\
 \Phi_r&=\frac{aP}{\Delta},
 &\Phi_\theta&=L_z\csc^2\theta-aE.
 \label{eq:Tphi_parts}
\end{align}

Using Mino time $\lambda$, the $r$ and $\theta$ coordinate motions can be separated each other. To describe the geometric nature of orbits,  it is very useful to introduce two reparameterizations of $\psi$ and $\chi$ , which are defined by
\begin{equation}
r = \frac{pM}{1 + e\cos\psi}\;,\qquad
\cos\theta = \sqrt{1-x^2}\cos(\chi + \chi_0)\;,
\label{eq:rdefthdef}
\end{equation}
where $x=\cos[\pi/2-{\rm sgn}(L_z)\theta_m]$ is a useful parameter to describe inclination: $x$ is changing smoothly from 1 to -1 as orbits vary from prograde equatorial to retrograde equatorial, with $L_z$ having the same sign as $x$. These transformations replace the variables $r$ and $\theta$ with
secularly accumulating angles $\psi$ and $\chi$. $\theta_m$ is the minimum value of polar motion, and $\theta_{\rm inc} = \pi/2-\theta_m$ describes inclination of orbit relative to the equatorial plane, determining the changing of the integrals of motion $Q$.
As $\psi$ and $\chi$ evolve from 0 to $2\pi$, $r$ and $\theta$ move through their full ranges of motion.  We define $\chi = \psi = 0$ at $\lambda = 0$.
In fact,  we choose $\psi_0 = 0$, which amounts to setting $\lambda =0$ at a moment that the orbit passes through periapsis and orbit passes apoapsis  when $\psi=\pi$.  The offset phase $\chi_0$ thus sets the value
of $\theta$ at periapsis.  Previous work (e.g., {\cite{Drasco:2005kz}}) has
typically used $\chi_0 = 0$ as well.  Following this reference, $\chi_0 = 0$ will label the ``fiducial geodesic.''  We will use it as a reference geodesic for some of the calculations in following Sec.\ {\ref{sec:electromagnetic:modes}}.

In generic equations for Kerr BH, Eqs.\ \eqref{eq:geodesic:R}--\eqref{eq:geodesic:phi}, Kerr orbits can be parameterized by three conserved constants $E$, $L_z$ and $Q$.  The reparameterization
(\ref{eq:rdefthdef}) maps those constants to parameters that describe
the orbit's geometry parameters: semi-latus rectum $p$, eccentricity
$e$, and minimum angle $\theta_m$.  These quantities are also
conserved along a geodesic orbit.  Schmidt {\cite{Schmidt:2002qk}} has given the
closed-form expressions for converting between $(E,L_z,Q)$ and
$(p,e,\theta_m)$, the parameters $(E,L_z,Q)$ or $(p,e,\theta_m)$ completely specify a geodesic for
our purposes here.

Observing that the radial and polar geodesic equations in Eqs.\eqref{eq:geodesic:R} and \eqref{eq:geodesic:Theta} have two periods, $\Lambda_r$ and $\Lambda_\theta$, which corresponds to two orbital fundamental frequencies, $\Upsilon_r$ and $\Upsilon_\theta$, determined by
\begin{equation}
 \Upsilon_r=\frac{2\pi}{\Lambda_r},
 \qquad
 \Upsilon_\theta=\frac{2\pi}{\Lambda_\theta}.
\end{equation}
The secular evolution of $t$ and $\phi$ is characterized by the Mino-time averages $\Gamma\equiv\langle dt/d\lambda\rangle$ and $\Upsilon_\phi\equiv\langle d\phi/d\lambda\rangle$, which can be expressed as
\begin{align}
 \Gamma&=\avg{T_r}_r+\avg{T_\theta}_\theta,\\
 \Upsilon_\phi&=\avg{\Phi_r}_r+\avg{\Phi_\theta}_\theta,
\end{align}
Here $\langle\cdot\rangle_r$ and $\langle\cdot\rangle_\theta$ denote averages over one radial and one polar Mino-time period, respectively. Separating the secular and oscillatory contributions gives
\begin{align}
 t(\lambda)&=\Gamma\lambda+\Delta t_r(\lambda)+\Delta t_\theta(\lambda),\\
 \phi(\lambda)&=\Upsilon_\phi\lambda+\Delta\phi_r(\lambda)+\Delta\phi_\theta(\lambda).
 \label{eq:tphi_decomp}
\end{align}
Here the $\Delta t_A$ and $\Delta\phi_A$ vary periodically at the harmonics of $\Upsilon_r$ and $\Upsilon_\theta$, please refer to Ref~\cite{Drasco:2003ky} for detailed discussion. The coordinate-time frequencies are
\begin{equation}
 \Omega_i=\frac{\Upsilon_i}{\Gamma},
 \qquad
 i=r,\theta,\phi,
\end{equation}
and the discrete frequency spectrum is
\begin{equation}
\omega_{mkn}=m\Omega_\phi+k\Omega_\theta+n\Omega_r
 =\frac{m\Upsilon_\phi+k\Upsilon_\theta+n\Upsilon_r}{\Gamma}.
 \label{eq:omega}
\end{equation}
We assume that the generic orbit is nonresonant,
$\Upsilon_\theta/\Upsilon_r\notin\mathbb{Q}$.The three fundamental frequencies can be computed either from the constants of motion $(E,L_z,Q)$ or from the equivalent geometric orbital parameters $(p,e,\theta_m)$.

\section{Derivation of Carter flux influenced by electromagnetic radiation}
\label{sec:scalarization}
In this section, we introduce the electromagnetic radiation in detail from strong-field generic orbits in Kerr spacetime. Since some papers have introduced the derivation of electromagnetic energy and angular momentum fluxes in Teukolsky formula, so we present a summary in subsection~\ref{sec:fluxes}, the derivation of Carter fluxes in subsections~\ref{sec:carter:rate:derivation} and subsection~\ref{sec:electromagnetic:modes}.

\subsection{Energy and angular-momentum fluxes from electromagnetic Teukolsky variables}\label{sec:fluxes}
We begin with the Maxwell equations in the Kerr background,
\begin{equation}
\nabla_{\nu}F^{\mu\nu}=4\pi J^\mu,
\qquad
\nabla_{[\mu}F_{\nu\sigma]}=0,
\label{eq:Maxwell}
\end{equation}
where $F_{\mu\nu}$ is the Faraday tensor.  We write the physical electric
charge of the secondary as $q/m_p$, so that $q$ is the charge-to-mass
parameter used throughout this work.  The corresponding point-particle
current is
\begin{equation}
J^\mu(x)
=
q_v\,m_p
\int
u^\mu(\tau)
\frac{\delta^{(4)}[x-z(\tau)]}{\sqrt{-g}}
\,d\tau ,
\label{eq:Jmu}
\end{equation}
where $z^\mu(\tau)$ is the particle worldline and
$u^\mu=dz^\mu/d\tau$.

Teukolsky showed that the extreme Newman--Penrose Maxwell scalars satisfy decoupled and separable equations on the Kerr spacetime~\cite{Teukolsky:1972my,Teukolsky:1973ha}.  For the spin-$+1$ scalar
\begin{equation}
\phi_0
\equiv
F_{\mu\nu}l_+^\mu m_+^\nu ,
\end{equation}
we use the compact covariant form~\cite{Bini:2002jx}
\begin{equation}
\left[
\left(\nabla_\mu+\Gamma_\mu\right)
\left(\nabla^\mu+\Gamma^\mu\right)
-4\psi_2
\right]\phi_0
=
4\pi J_0 ,
\label{eq:4Dteuk}
\end{equation}
where $\Gamma^\mu$ is the connection vector defined in
Ref.~\cite{Bini:2002jx}, $\psi_2$ is the nonvanishing background Weyl
scalar in the corresponding tetrad convention, and $J_0$ is obtained by
applying the spin-$+1$ Teukolsky source operator to the tetrad projections
of $J^\mu$.

We separate the field with the following equation
\begin{equation}
\phi_0
=
\int d\omega
\sum_{\ell m}
\Delta^{-1}
R_{+1}^{\ell m\omega}(r)
S_{+1}^{\ell m\omega}(\theta)
e^{-i\omega t+im\phi},
\label{eq:phi0-expansion}
\end{equation}
which gives
\begin{align}
\left(
\Delta\mathcal D^\dagger\mathcal D
+2i\omega r
-\lambda_{\ell m\omega}
\right)
R_{+1}^{\ell m\omega}(r)
&=
\mathcal T_{+1}^{\ell m\omega}(r),
\label{eq:radial-sourced}
\\
\left[
\mathcal L^\dagger
\left(\mathcal L+\cot\theta\right)
-2a\omega\cos\theta
+\lambda_{\ell m\omega}
\right]
S_{+1}^{\ell m\omega}(\theta)
&=
0.
\label{eq:angular-sourced}
\end{align}
Here $\lambda_{\ell m\omega}$ is the angular separation constant, and the
mode labels will be suppressed when no confusion can arise.  We define
\begin{subequations}
\begin{align}
\mathcal D
&=
\partial_r-\frac{i\mathcal K}{\Delta},
&
\mathcal D^\dagger
&=
\partial_r+\frac{i\mathcal K}{\Delta},
\\
\mathcal L
&=
\partial_\theta+\mathcal Q,
&
\mathcal L^\dagger
&=
\partial_\theta-\mathcal Q,
\end{align}
\label{eq:DLoperators}
\end{subequations}
with
\begin{equation}
\mathcal K
=
\omega(r^2+a^2)-am,
\qquad
\mathcal Q
=
m\csc\theta-a\omega\sin\theta.
\label{eq:KQoperators}
\end{equation}
The angular eigenvalues and spin-weighted spheroidal harmonics may be
computed numerically using, for example, the Black Hole Perturbation
Toolkit~\cite{BHPToolkit}.

For a generic inclined and eccentric orbit, the delta
distribution in Eq.~\eqref{eq:Jmu} is rewritten as
\begin{align}
\frac{\delta^{(4)}[x-z(\tau)]}{\sqrt{-g}}
={}&
\frac{1}{\Sigma\sin\theta}
\delta[t-t_p(\tau)]
\delta[r-r_p(\tau)]
\nonumber\\
&\times
\delta[\theta-\theta_p(\tau)]
\delta[\phi-\phi_p(\tau)].
\label{eq:delta4_generic}
\end{align}
It is important that the polar delta function in
Eq.~\eqref{eq:delta4_generic} is $\delta[\theta-\theta_p(\tau)]$. 
Using the relationship of delta function
\begingroup\footnotesize
\begin{equation}
\delta(t-t_p)
=
\frac{1}{2\pi}
\int d\omega\,
e^{-i\omega(t-t_p)},
\quad
\delta(\phi-\phi_p)
=
\frac{1}{2\pi}
\sum_m
e^{im(\phi-\phi_p)},
\label{eq:delta_tf}
\end{equation}
\endgroup
and the spheroidal harmonics 
\begin{align}
\delta(\theta - \theta_0) &= 2 \pi \sum_\ell S_{+1}^{\ell m \gamma}(\theta) S_{+1}^{\ell m \gamma}(\theta_0) , \\
\frac{\partial}{\partial \theta} \delta(\theta - \theta_0) &= - 2 \pi \sum_\ell S_{+1}^{\ell m \gamma}(\theta) S_{+1}^{\ell m \gamma \prime}(\theta_0) \;,
\end{align}
we can obtain expression for the radial source function in  Eq.~\eqref{eq:radial-sourced}. Its generic structure may be written as
\begingroup\small
\begin{align}
\mathcal{T}_{+1} = \sqrt{2} q_v \Delta \int& \frac{d \tau}{r_0}  e^{- i \chi_0}  \Bigg\{  \Big( A(r_0) S_{+1}(\theta_0) - u_{l^+}^{(0)} 
S_{+1}^\prime(\theta_0) \Big) 
\nonumber \\ &\times
\delta(r - r_0) - u_{m^+}^{(0)} S_{+1}(\theta_0) \delta^\prime(r - r_0)  \Bigg\} \;,\label{eq:source:vector}
\end{align}
\endgroup
where $A(r_0)$ is defined by 
\begin{align}
A_{\pm}(r_0) &= u_{l^\pm}^{(0)} \left( \mp \mathcal{Q}+ \frac{i a}{r_0} \right) - u^{(0)}_{m^{\pm}} \left(\mp \frac{i \cal K }{\Delta} + \frac{1}{r_0} \right) , \label{eq:Adef}
\end{align}
where 
\begin{equation}
\begin{aligned}
u_{l^\pm}^{(0)} &= l_{\pm}^\mu u_\mu = \frac{r_0^2 \dot{r}_0 \mp (r_0^2 + a^2) E \pm a L}{\Delta_0} , \\
u_{m^\pm}^{(0)} &= m_{\pm}^\mu u_\mu = \pm i (L - a E) .
\end{aligned} 
\label{u0def}
\end{equation}
They depend on the instantaneous generic worldline through $r_p$, $\theta_p$ and the tetrad projections of $u^\mu$.

To compute the electromagnetic fluxes, we examine the asymptotic behavior of
the homogeneous radial solutions near the event horizon and at spatial
infinity.  The solution $R_{\ell m\omega}^{\infty}(r)$ is chosen to satisfy
a purely outgoing boundary condition as $r\rightarrow\infty$, whereas
$R_{\ell m\omega}^{H}(r)$ satisfies a purely ingoing boundary condition as
$r\rightarrow r_+$.  Using these two independent homogeneous solutions, the
inhomogeneous radial function is constructed through the Green-function
representation and takes the asymptotic form
\begin{equation}
R_{\ell m\omega}(r)
=
\begin{cases}
Z^H_{\ell m\omega}(\chi_0)\,
R_{\ell m\omega}^{\infty}(r),
&
r\rightarrow\infty,
\\[1mm]
Z^\infty_{\ell m\omega}(\chi_0)\,
R_{\ell m\omega}^{H}(r),
&
r\rightarrow r_+,
\end{cases}
\label{eq:Rasymptotic}
\end{equation}
where $\chi_0$ denotes the relative initial phase between the radial and
polar motions.  With a fixed normalization of the homogeneous solutions,
the corresponding Green-function amplitudes can be written as
\begingroup\footnotesize
\begin{equation}
Z^\star_{\ell m\omega}(\chi_0)
=
C^\star_{\ell m\omega}
\int_{r_+}^{\infty}
dr'\,
\frac{
R^\star_{\ell m\omega}(r')
\mathcal T_{\ell m\omega}(r',\chi_0)
}{
\Delta(r')^2
},
\quad
\star\in\{H,\infty\},
\label{eq:Z1}
\end{equation}
\endgroup
where $C^\star_{\ell m\omega}$ contains the Wronskian and the normalization
constants associated with the chosen homogeneous solutions; see
Ref.~\cite{German:2023bye} for the corresponding spin-$1$ construction.

After the radial integration over the delta functions in
Eq.~\eqref{eq:source:vector}, the amplitude becomes a Fourier transform along
the worldline,
\begingroup\small
\begin{align}
Z^\star_{\ell m\omega}(\chi_0)
={}&
C^\star_{\ell m\omega}
\int_{-\infty}^{\infty}
dt\,
e^{i[\omega t-m\phi_p(t)]}
I^\star_{\ell m\omega}
[r_p(t),\theta_p(t,\chi_0)]
\nonumber\\
={}&
C^\star_{\ell m\omega}
\int_{-\infty}^{\infty}
d\lambda\,
e^{i(\omega\Gamma-m\Upsilon_\phi)\lambda}
J^\star_{\ell m\omega}
[r_p(\lambda),\theta_p(\lambda,\chi_0)].
\label{eq:Z2}
\end{align}
\endgroup
Here
\begin{align}
J^\star_{\ell m\omega}(r_p,\theta_p)
={}&
I^\star_{\ell m\omega}(r_p,\theta_p)
\frac{dt}{d\lambda}
\nonumber\\
&\times
e^{i[\omega\Delta t(r_p,\theta_p)
-m\Delta\phi(r_p,\theta_p)]},
\label{eq:Jdef}
\end{align}
where
\begin{align}
t_p(\lambda)
&=
\Gamma\lambda
+
\Delta t_r(\lambda)
+
\Delta t_\theta(\lambda),
\nonumber\\
\phi_p(\lambda)
&=
\Upsilon_\phi\lambda
+
\Delta\phi_r(\lambda)
+
\Delta\phi_\theta(\lambda).
\label{eq:tphi_source}
\end{align}
The functions $\Delta t_{\theta,r}$ and $\Delta\phi_{\theta,r}$ are periodic in the radial and polar Mino-time phases~\cite{Drasco:2003ky,Drasco:2005kz}.

Because Mino time separates the radial and polar motions, the periodic
source function admits a two-dimensional Fourier decomposition,
\begin{equation}
J^\star_{\ell m\omega}
=
\sum_{kn}
J^\star_{\omega\ell mkn}(\chi_0)
e^{-i(k\Upsilon_\theta+n\Upsilon_r)\lambda},
\label{eq:Jdecomp}
\end{equation}
where
\begingroup\small
\begin{align}
J^\star_{\omega\ell mkn}(\chi_0)
={}&
\frac{\Upsilon_r\Upsilon_\theta}{(2\pi)^2}
\int_0^{2\pi/\Upsilon_\theta}
d\lambda^\theta
\int_0^{2\pi/\Upsilon_r}
d\lambda^r
\nonumber\\
&\times
e^{i(k\Upsilon_\theta\lambda^\theta
+n\Upsilon_r\lambda^r)}
J^\star_{\ell m\omega}
[
r_p(\lambda^r),
\theta_p(\lambda^\theta,\chi_0)
].
\label{eq:Jcoef}
\end{align}
\endgroup
Combining Eqs.~\eqref{eq:Z2}--\eqref{eq:Jcoef} gives the discrete
frequency spectrum
\begin{equation}
Z^\star_{\ell m\omega}(\chi_0)
=
\sum_{kn}
Z^\star_{\ell mkn}(\chi_0)
\delta(\omega-\omega_{mkn}),
\label{eq:Zexpand}
\end{equation}
with
\begin{equation}
\omega_{mkn}
=
m\Omega_\phi+k\Omega_\theta+n\Omega_r
=
\frac{
m\Upsilon_\phi+k\Upsilon_\theta+n\Upsilon_r
}{
\Gamma
},
\label{eq:omega_source}
\end{equation}
and
\begingroup\small
\begin{align}
Z^\star_{\ell mkn}(\chi_0)
={}&
\frac{
2\pi C^\star_{\ell m\omega_{mkn}}
}{
\Gamma
}
J^\star_{\omega_{mkn}\ell mkn}(\chi_0)
\nonumber\\
={}&
\frac{
C^\star_{\ell m\omega_{mkn}}
\Upsilon_r\Upsilon_\theta
}{
2\pi\Gamma
}
\int_0^{2\pi/\Upsilon_\theta}
d\lambda^\theta
\int_0^{2\pi/\Upsilon_r}
d\lambda^r
\nonumber\\
&\times
e^{i(k\Upsilon_\theta\lambda^\theta
+n\Upsilon_r\lambda^r)}
J^\star_{\ell m\omega_{mkn}}
[
r_p(\lambda^r),
\theta_p(\lambda^\theta,\chi_0)
].
\label{eq:Zcoef}
\end{align}
\endgroup
The factor $C^\star_{\ell m\omega}$ in Eq.~\eqref{eq:Zcoef} is essential:
it follows directly from the Green-function normalization in
Eq.~\eqref{eq:Z1} and must not be dropped unless it has first been absorbed
into the definition of $J^\star_{\ell m\omega}$. For a nonresonant orbit, changing the initial radial--polar phase affects a discrete amplitude only through a phase.  Defining the fiducial amplitude 
\begin{equation}
\check Z^\star_{\ell mkn}
\equiv
Z^\star_{\ell mkn}(\chi_0=0),
\end{equation}
one may write~\cite{Drasco:2003ky}
\begin{equation}
Z^\star_{\ell mkn}(\chi_0)
=
e^{i\xi_{mkn}(\chi_0)}
\check Z^\star_{\ell mkn}.
\label{eq:Zphaseshift}
\end{equation}
Consequently,
\begin{equation}
\left|
Z^\star_{\ell mkn}(\chi_0)
\right|^2
=
\left|
\check Z^\star_{\ell mkn}
\right|^2,
\end{equation}
so the nonresonant dissipative flux on each mode is independent of the
choice of fiducial phase.

Finally, the physical electromagnetic energy and axial angular-momentum
fluxes are most cleanly expressed in terms of asymptotic spin-$1$ Maxwell
amplitudes.  Denoting by
$Z_{\ell mkn}^{\infty}$ the outgoing amplitude at infinity and by
$Z_{\ell mkn}^{H}$ the ingoing amplitude at the horizon,  the energy fluxes are given by
\begin{equation}
\dot{\mathcal E}_{\ell mkn}^{v,\infty}\equiv 
\left\langle \frac{dE^{v,\infty}}{dt}\right\rangle
=\frac{1}{8\pi}\left|Z_{\ell mkn}^{v,\infty}\right|^2,
\label{eq:EMfluxInf}
\end{equation}
and
\begingroup\small
\begin{equation}
\dot{\mathcal E}_{\ell mkn}^{v,H} \equiv 
\left\langle \frac{dE^{v,H}}{dt}\right\rangle
=\frac{1}{8\pi}
\frac{\omega_{mkn}}{2Mr_+
\left(\omega_{mkn}-m\Omega_H\right)}
\left|Z_{\ell mkn}^{v,H}\right|^2.
\label{eq:EMfluxH}
\end{equation}
\endgroup
The corresponding angular-momentum flux satisfies
\begin{equation}
\dot{\mathcal L}_{\ell mkn}^{v,\star}
=
\frac{m}{\omega_{mkn}}
\dot{\mathcal E}_{v,\ell mkn}^{\star},
\qquad
\star\in\{\infty,H\}.
\label{eq:EMLflux}
\end{equation}
The total electromagnetic energy and angular-momentum fluxes are obtained by summing over all multipole harmonics,
\begin{equation}
\begin{aligned}  
\dot{\cal E}_v &= \dot{\mathcal{E}}^{H}_v + \dot{\mathcal{E}}^{\infty}_v\;,\\
\dot{\cal L}_v &= \dot{\mathcal{L}}^{H}_v + \dot{\mathcal{L}}^{\infty}_v\;,\\
\dot{\mathcal E}^{\star}_v, \dot{\mathcal L}^{\star}_v
&=\sum_{\ell mkn}
\dot{\mathcal{E}}^{\star}_{v,\ell mkn}, \dot{\cal{L}}_{v,\ell mkn}^{\star}
\quad
\star\in\{\infty,H\}.
\label{eq:EMtotalflux}
\end{aligned}
\end{equation}
Here we have obtained the energy and angular-momentum fluxes radiated by electromagnetic perturbation, which are particularly important in the Carter flux derivation in the below section. 


\subsection{Derivation of the Carter-constant evolution}
\label{sec:carter:rate:derivation}

For the spin-weight $s=-1$ electromagnetic perturbation, the radiative field
strength is constructed from the radiative vector potential as
\begin{equation}
F^{\rm rad}_{\alpha\beta}
=
\nabla_\alpha A^{\rm rad}_\beta
-
\nabla_\beta A^{\rm rad}_\alpha ,
\label{eq:FfromA}
\end{equation}
and we introduce
\begin{equation}
P^\beta
\equiv
K^{\alpha\beta} u_\alpha .
\label{eq:Pdef}
\end{equation}
The Carter constant,
$ K=K_{\alpha\beta}u^\alpha u^\beta$,
is conserved along an unperturbed Kerr geodesic.  In the presence of the
radiative electromagnetic self-force, its evolution along the worldline is
obtained by differentiating with respect to the proper time $\tau$:
\begin{align}
\frac{d K}{d\tau}
&=
u^\rho\nabla_\rho
\left(
K_{\alpha\beta}u^\alpha u^\beta
\right)
\nonumber\\
&=
u^\rho u^\alpha u^\beta
\nabla_\rho K_{\alpha\beta}
+
2K_{\alpha\beta}u^\alpha
u^\rho\nabla_\rho u^\beta.
\end{align}
The first term vanishes by virtue of the Killing-tensor equation,
$\nabla_{(\gamma}K_{\alpha\beta)}=0$.  Using Eq.~\eqref{eq:Pdef}, the
evolution of $K$ therefore becomes
\begin{equation}
\frac{d K}{d\tau}
=2 K_{\alpha\beta}u^\alpha
F_{\rm rad}^{\beta\gamma}u_\gamma
=
2P_\beta F^{\rm rad}_{\beta\gamma}u^\gamma .
\label{eq:localK}
\end{equation}

We next simplify the right-hand side of Eq.~\eqref{eq:localK}.  Substituting Eq.~\eqref{eq:FfromA}, one finds
\begin{align}
P^\beta
F_{\beta\gamma}^{\rm rad}
u^\gamma
&=P^\beta u^\gamma
\nabla_\beta A_\gamma^{\rm rad}
-P^\beta u^\gamma
\nabla_\gamma A_\beta^{\rm rad}.\label{eq:eq28step1}
\end{align}
For the first term, applying the product rule to the scalar
$ u^\gamma A_\gamma^{\rm rad}$ gives
\begin{equation}
\nabla_\beta
\left( u^\gamma A_\gamma^{\rm rad}\right)
=
(\nabla_\beta u^\gamma)
A_\gamma^{\rm rad}
+ u^\gamma \nabla_\beta A_\gamma^{\rm rad},
\end{equation}
and hence
\begin{equation}
P^\beta u^\gamma
\nabla_\beta A_\gamma^{\rm rad}
=
P^\beta\nabla_\beta
\left( u^\gamma A_\gamma^{\rm rad}\right)
-P^\beta A_\gamma^{\rm rad}
\nabla_\beta u^\gamma .
\label{eq:eq28firstterm}
\end{equation}
For the second term in Eq.~\eqref{eq:eq28step1}, differentiating the scalar
$P^\beta A_\beta^{\rm rad}$ along the worldline yields
\begin{align}
\frac{d}{d\tau}
\left(
P^\beta A_\beta^{\rm rad}
\right)
&\equiv u^\gamma\nabla_\gamma
\left(
P^\beta A_\beta^{\rm rad}
\right)
\nonumber\\
&=
A_\beta^{\rm rad}
 u^\gamma\nabla_\gamma P^\beta
+P^\beta u^\gamma
\nabla_\gamma A_\beta^{\rm rad}
\nonumber\\
&=
A_\beta^{\rm rad}
\frac{DP^\beta}{d\tau}
+ P^\beta u^\gamma
\nabla_\gamma A_\beta^{\rm rad}.
\label{eq:eq28secondproduct}
\end{align}
Therefore,
\begin{equation}
-
P^\beta u^\gamma
\nabla_\gamma A_\beta^{\rm rad}
=-\frac{d}{d\tau}
\left(
P^\beta A_\beta^{\rm rad}
\right)
+
A_\beta^{\rm rad}
\frac{DP^\beta}{d\tau}.
\label{eq:eq28secondterm}
\end{equation}
Combining Eqs.~\eqref{eq:eq28firstterm} and
\eqref{eq:eq28secondterm}, Eq.~\eqref{eq:eq28step1} can be written as
\begin{equation}
\begin{aligned}
P^\beta F_{\beta\gamma}^{\rm rad} u^\gamma
={}&
P^\beta\nabla_\beta
\left(
 u^\gamma A_\gamma^{\rm rad}
\right)
-
P^\beta A_\gamma^{\rm rad}
\nabla_\beta u^\gamma
\\
&-
\frac{d}{d\tau}
\left(
P^\beta A_\beta^{\rm rad}
\right)
+
A_\beta^{\rm rad}
\frac{DP^\beta}{d\tau}.
\end{aligned}
\label{eq:PFu:four:terms}
\end{equation}
Equation~\eqref{eq:PFu:four:terms} is an exact tensor identity.  To proceed,
we evaluate the transport of $P_\mu$ along the geodesic:
\begin{align}
\frac{DP_\mu}{d\tau}
&=
 u^\rho\nabla_\rho
\left(
K_{\mu\nu} u^\nu
\right)
\nonumber\\
&=
 u^\rho u^\nu
\nabla_\rho K_{\mu\nu}
+
K_{\mu\nu}
 u^\rho\nabla_\rho u^\nu .
\end{align}
The second term vanishes for the background geodesic,
$ u^\rho\nabla_\rho u^\nu=0$.
Using the Killing-tensor identity then gives
\begin{equation}
\frac{DP_\mu}{d\tau}
=
-\frac12
 u^\rho u^\nu
\nabla_\mu K_{\rho\nu}.
\label{eq:Ptransport2}
\end{equation}

Expanding
$\nabla_\mu(K_{\rho\nu} u^\rho u^\nu)=0$
yields
\begin{equation}
 u^\rho u^\nu
\nabla_\mu K_{\rho\nu}
+
2K_{\rho\nu}
u^\rho\nabla_\mu u^\nu
=
0,
\label{eq:Kderivative}
\end{equation}
which, together with Eq.~\eqref{eq:Ptransport2}, implies
\begin{equation}
\frac{DP^\beta}{d\tau}
=P^\gamma\nabla_\gamma u^\beta.
\label{eq:Ptransport}
\end{equation}
Consequently, the second and fourth terms on the right-hand side of
Eq.~\eqref{eq:PFu:four:terms} cancel, and we obtain
\begin{equation}
P^\beta
F_{\beta\gamma}^{\rm rad}
 u^\gamma
=
P^\beta\nabla_\beta
\left(
u^\gamma A_\gamma^{\rm rad}
\right)
-
\frac{d}{d\tau}
\left(
P^\beta A_\beta^{\rm rad}
\right).
\label{eq:keyidentity}
\end{equation}
Using Eq.~\eqref{eq:keyidentity} and the relation
$d\tau=\Sigma\,d\lambda$, the evolution of the Carter constant in Mino time
can be written as
\begin{equation}
\frac{dK}{d\lambda}
=
2\Sigma
P^\beta\nabla_\beta
\left(
 u^\gamma A_\gamma^{\rm rad}
\right)
-
2
\frac{d}{d\lambda}
\left(
P^\beta A_\beta^{\rm rad}
\right).
\label{eq:KlambdaPrePsi}
\end{equation}
Introducing the auxiliary scalar
\begin{equation}
\Psi_V^{\rm rad}(x)
\equiv
\Sigma(x)
u^\gamma(x)
A_\gamma^{\rm rad}(x),
\label{eq:PsiV}
\end{equation}
and using $P^\beta=K^{\alpha\beta} u_\alpha$, we obtain
\begin{equation}
\frac{d K}{d\lambda}
=
2\Sigma
K^{\alpha\beta} u_\alpha
\partial_\beta
\left(
\frac{\Psi_V^{\rm rad}}{\Sigma}
\right)
-
2\frac{d}{d\lambda}
\left(
P^\beta A_\beta^{\rm rad}
\right).
\label{eq:KlambdaWithTD}
\end{equation}

For a bound orbit, $P^\beta A_\beta^{\rm rad}$ is a bounded scalar along the
worldline. Therefore, the long-time average of the total derivative in
Eq.~\eqref{eq:KlambdaWithTD} vanishes, and the equation of change rate of $dK/d\lambda$ reduces to
\begin{equation}
\left\langle
\frac{d K}{d\lambda}
\right\rangle
=\left\langle
\left.
2\Sigma K^{\alpha\beta} u_\alpha
\partial_\beta
\left(
\frac{\Psi_V^{\rm rad}}{\Sigma}
\right)
\right|_{x\rightarrow z(\lambda)}
\right\rangle .
\label{eq:VB5}
\end{equation}
Here $x\rightarrow z(\lambda)$ denotes evaluation on the geodesic worldline
$z(\lambda)$, and the long-time average is defined by
\begin{equation}
\left\langle
\frac{dK}{d\lambda}
\right\rangle
\equiv
\lim_{L\rightarrow\infty}
\frac{1}{2L}
\int_{-L}^{L}
d\lambda\,
\frac{dK}{d\lambda}.
\label{eq:VB6}
\end{equation}
Equations~\eqref{eq:VB5} and \eqref{eq:VB6} constitute the direct
electromagnetic analogue of the corresponding radiative-field expression
derived in Ref.~\cite{Flanagan:2012kg}.  Since the derivation starts from
the gauge-invariant Faraday tensor, the orbit-averaged result is invariant
under regular gauge transformations.

We next simplify Eq.~\eqref{eq:VB5} by using the explicit form of the Kerr
Killing tensor, Eq.~\eqref{eq:KillingTensor}, together with
$\Sigma u^\alpha=dx^\alpha/d\lambda$.  Evaluated on the worldline, the
quantity entering the Carter flux becomes
\begin{align}
&
2\Sigma K^{\alpha\beta} u_\alpha
\partial_\beta
\left(
\frac{\Psi_V}{\Sigma}
\right)
\nonumber\\
={}&
2\Sigma\left[
(L_z-aE\sin^2\theta)
(\csc^2\theta\,\partial_\phi+a\partial_t)
+
\frac{d\theta}{d\lambda}\partial_\theta
\right]
\left(
\frac{\Psi_V}{\Sigma}
\right)
\nonumber\\
&-
2a^2\cos^2\theta
\frac{d}{d\lambda}
\left(
\frac{\Psi_V}{\Sigma}
\right).
\label{eq:VB12}
\end{align}
The terms involving the polar motion can be rearranged as
\begin{align}
&
2\Sigma\frac{d\theta}{d\lambda}\partial_\theta
\left(
\frac{\Psi_V}{\Sigma}
\right)
-
2a^2\cos^2\theta
\frac{d}{d\lambda}
\left(
\frac{\Psi_V}{\Sigma}
\right)
\nonumber\\
={}&
2\frac{d\theta}{d\lambda}\partial_\theta\Psi_V
-
2\frac{\Psi_V}{\Sigma}
\frac{d\theta}{d\lambda}\partial_\theta\Sigma
-
2a^2
\frac{d}{d\lambda}
\left(
\cos^2\theta\frac{\Psi_V}{\Sigma}
\right)
\nonumber\\
&+
2a^2\frac{\Psi_V}{\Sigma}
\frac{d}{d\lambda}\cos^2\theta.
\label{eq:VB13}
\end{align}
The third term in Eq.~\eqref{eq:VB13} is a total derivative and vanishes
after orbit averaging.  Using
\begin{equation}
\partial_\theta\Sigma
=
a^2\partial_\theta\cos^2\theta,
\qquad
\frac{d}{d\lambda}\cos^2\theta
=
\frac{d\theta}{d\lambda}
\partial_\theta\cos^2\theta,
\end{equation}
the second and fourth terms cancel.  We therefore obtain
\begin{align}
\left\langle
\frac{d K}{d\lambda}
\right\rangle
=
2\Bigg\langle
\Bigg[
&(L_z\csc^2\theta-aE)\partial_\phi
+a(L_z-aE\sin^2\theta)\partial_t
\nonumber \\ &  +
\frac{d\theta}{d\lambda}\partial_\theta
\Bigg] \Psi_V^{\rm rad}\Big|_{x\rightarrow z(\lambda)}
\Bigg\rangle .
\label{eq:VB15}
\end{align}
Equation~\eqref{eq:VB15} is the vector analogue of Eq.~(B15) of
Ref.~\cite{Flanagan:2012kg}.  Importantly, the reduction from
Eq.~\eqref{eq:VB5} to Eq.~\eqref{eq:VB15} depends only on the Kerr
geodesic structure and the Killing tensor; no additional assumption about
the spin of the radiative field enters this geometric simplification.

\subsection{Mode decomposition of the electromagnetic perturbation}
\label{sec:electromagnetic:modes}

Having expressed the change rate of the constant $K$ in terms of
the radiative electromagnetic field, we now perform a mode decomposition
to obtain explicit flux formulas.  Following the general strategy of
Appendix~B.3 of Ref.~\cite{Flanagan:2012kg}, the radiative field is
decomposed into ``out'' and ``down'' components,
\begin{equation}
\Psi_V^{\rm rad}
=
\Psi_V^{\rm out}
+
\Psi_V^{\rm down}.
\label{eq:outdown}
\end{equation}
These components are constructed from the mode functions
$\Phi_{\ell m\omega}$ with two following equations
\begingroup\small
\begin{align}
\Psi_V^{\rm out}
&=
\int d\omega
\sum_{\ell m}
\frac{1}{8i\omega}
\left[
\Phi_{\ell m\omega}^{\infty}
\int d\lambda'\,
\overline{\Phi}_{\ell m\omega}^{\infty}
\right]
+\mathrm{c.c.},
\label{eq:PsiV_inf}
\\
\Psi_V^{\rm down}
&=
\int d\omega
\sum_{\ell m}
\frac{1}
{16iMr_+p_m}
\left[
\Phi_{\ell m\omega}^{H}
\int d\lambda'\,
\overline{\Phi}_{\ell m\omega}^{H}
\right]
+\mathrm{c.c.},
\label{eq:PsiV_H}
\end{align}
\endgroup
where $p_m=\omega-m\Omega_H$. The notation ``c.c.'' denotes the complex-conjugate contribution.

We first consider the ``out'' sector.  Evaluating $\Phi_{\ell m\omega}^{\rm out}$ on the geodesic worldline gives
\begin{equation}
\Phi^{\rm out}_{\ell m\omega}[z(\lambda)]
=
\overline J^H_{\ell m\omega}(\lambda)
e^{-i\lambda(\Gamma\omega-m\Upsilon_\phi)}.
\label{eq:modefunc_fourier}
\end{equation}
The quasiperiodic source can then be decomposed into discrete harmonics,
leading to
\begingroup\small
\begin{equation}
\Psi_V^{\rm out}(x)
=
\int d\omega
\left[
\sum_{\ell mkn}
\frac{
Z^H_{\ell mkn}
\delta(\omega-\omega_{mkn})
}{
8i\omega
}
\Phi_{\ell m\omega}^{\rm out}(x)
\right]
+
{\rm c.c.}
\label{eq:Psi_out_rad2}
\end{equation}
\endgroup
Substituting this expression into Eq.~\eqref{eq:VB15}, the contribution to
the Carter flux at infinity becomes
\begin{align}
\left\langle
\frac{dK^\infty}{d\lambda}
\right\rangle
=\Bigg\langle
\sum_{\ell mkn}
&\frac{Z^H_{\ell mkn}}{4i\omega_{mkn}}
\Bigg\{
\Bigg[
(\csc^2\theta L_z-a E)\partial_\phi
\nonumber  \\ &
+
a(L_z-a E\sin^2\theta)\partial_t
\nonumber  \\ & +
\frac{d\theta}{d\lambda}\partial_\theta
\Bigg]
\Phi_{\ell mkn}^{\rm out}
\Bigg\}
+
{\rm c.c.}
\Bigg\rangle .
\label{eq:dKdlambda2}
\end{align}
Here
$\Phi_{\ell mkn}^{\rm out}
\equiv
\Phi_{\ell m\omega_{mkn}}^{\rm out}$,
and the superscript $\infty$ indicates the contribution associated with
the outgoing sector.

For a generic bound Kerr geodesic, the radial and polar motions are
quasiperiodic in Mino time.  The coordinates $t$ and $\phi$ can be written
as
\begin{align}
t(\lambda)
&=
\Gamma\lambda
+
\Delta t_r(\lambda)
+
\Delta t_\theta(\lambda),
\nonumber\\
\phi(\lambda)
&=
\Upsilon_\phi\lambda
+
\Delta\phi_r(\lambda)
+
\Delta\phi_\theta(\lambda),
\end{align}
where the oscillatory pieces are periodic functions of the radial and polar
motions.  Following Ref.~\cite{Flanagan:2012kg}, the worldline mode
function may therefore be regarded as a biperiodic function of the radial
and polar Mino-time phases,
\begin{align}
\Phi^{\rm out}_{\ell mkn}
(\lambda^r,\lambda^\theta)
=&
f_{\ell mkn}[r(\lambda^r),\theta(\lambda^\theta)]
\nonumber\\
&\times
\exp
\Big\{
-ik\Upsilon_\theta\lambda^\theta
-in\Upsilon_r\lambda^r
\nonumber\\
&
-i\omega_{mkn}
[\Delta t_r(\lambda^r)
+\Delta t_\theta(\lambda^\theta)]
\nonumber\\
&+im
[\Delta\phi_r(\lambda^r)
+\Delta\phi_\theta(\lambda^\theta)]
\Big\}.
\label{Phiexplicit}
\end{align}
$f_{\ell mkn}[r(\lambda^r),\theta(\lambda^\theta)]$ is not an important function for our purposes. Applying the differential operator in Eq.~\eqref{eq:dKdlambda2} to the mode function in Eq.~\eqref{Phiexplicit} yields 
\begingroup\small
\begin{align}
\left\langle
\frac{dK^\infty}{d\lambda}
\right\rangle
&=
\Bigg\langle
\sum_{\ell mkn}
\frac{Z^H_{\ell mkn}}{4i\omega_{mkn}}
\Bigg\{
\Bigg[
\left(
\csc^2\theta L_z-aE
-
\frac{d\Delta\phi_\theta}{d\lambda^\theta}
\right)\partial_\phi
\nonumber\\
+
&
\left(
aL_z-a^2E\sin^2\theta
-
\frac{d\Delta t_\theta}{d\lambda^\theta}
\right)\partial_t
\nonumber\\
&+
ik\Upsilon_\theta
+\frac{d}{d\lambda^\theta}
\Bigg]
\Phi_{\ell mkn}^{\rm out}
\Bigg\}
+
{\rm c.c.}
\Bigg\rangle .
\label{eq:dKdlambda3}
\end{align}
\endgroup

Using the standard identities for generic Kerr geodesics
~\cite{Drasco:2005is},
\begin{align}
\csc^2\theta L_z-aE
-
\frac{d\Delta\phi_\theta}{d\lambda^\theta}
&=
\left\langle
\csc^2\theta L_z-aE
\right\rangle
\nonumber\\
&=
\langle\csc^2\theta\rangle L_z - a E,
\label{eq:ave1}
\\
a L_z-a^2E\sin^2\theta
-
\frac{d\Delta t_\theta}{d\lambda^\theta}
&=
\left\langle
aL_z-a^2E\sin^2\theta
\right\rangle
\nonumber\\
&=
aL_z
-
a^2E\langle\sin^2\theta\rangle,
\label{eq:ave2}
\end{align}
we obtain
\begin{align}
\left\langle
\frac{dK^\infty}{d\lambda}
\right\rangle
=\Bigg\langle
\sum_{\ell mkn}
\frac{Z^H_{\ell mkn}}{4i\omega_{mkn}}
\Bigg[ &i{\cal M}_{mkn} + ik\Upsilon_\theta
\nonumber  \\ &+
\frac{d}{d\lambda^\theta}
\Bigg]\Phi_{\ell mkn}^{\rm out} + {\rm c.c.}
\Bigg\rangle ,\label{eq:dKdlambda3a}
\end{align}
where
\begingroup\small
\begin{equation}
{\cal M}_{mkn}
=m \left(
\langle\csc^2\theta\rangle L_z-aE
\right)
-
a\omega_{mkn}
\left(
L_z-aE\langle\sin^2\theta\rangle
\right).
\label{eq:calMdef}
\end{equation}
\endgroup

The remaining mode dependence can be expanded as
\begin{align}
\Phi^{\rm out}_{\ell mkn}
(\lambda^r,\lambda^\theta)
=
\frac{\Gamma}{2\pi}
\sum_{\Delta n,\Delta k}
&
\overline Z^H_{
\omega_{mkn}\,\ell m\,k+\Delta k,\,n+\Delta n
}
\nonumber\\
&\times
e^{i\Delta k\Upsilon_\theta\lambda^\theta}
e^{i\Delta n\Upsilon_r\lambda^r}.
\label{eq:Phioutfull}
\end{align}
Substituting this expansion into Eq.~\eqref{eq:dKdlambda3a} gives
\begin{align}
\left\langle
\frac{dK^\infty}{d\lambda}
\right\rangle
=
\Bigg\langle
&\frac{\Gamma}{4\pi}
\sum_{\ell mkn}
\sum_{\Delta k,\Delta n}
\left[
{\cal M}_{mkn}
+
k\Upsilon_\theta
+
\Delta k\Upsilon_\theta
\right]
\nonumber\\
&\times
\frac{Z^H_{\ell mkn}}{8\pi\omega_{mkn}}
\overline Z^H_{\omega_{mkn}\,\ell m\,k+\Delta k,\,n+\Delta n}
\nonumber\\
&\times
e^{i\Delta k\Upsilon_\theta\lambda^\theta}
e^{i\Delta n\Upsilon_r\lambda^r}
+{\rm c.c.}\Bigg\rangle .\label{eq:dKdlambda3b}
\end{align}

In this work we restrict attention to nonresonant generic Kerr geodesics.
After identifying the radial and polar phase variables with the physical
Mino-time parameter, the long-time average of the off-diagonal Fourier
terms vanishes unless
$\Delta k\,\Upsilon_\theta+\Delta n\,\Upsilon_r=0$.  For a nonresonant
orbit,
\begin{align}
&
\lim_{L\rightarrow\infty}
\frac{1}{2L}
\int_{-L}^{L}
d\lambda\,
e^{i(\Delta k\Upsilon_\theta+\Delta n\Upsilon_r)\lambda}
\nonumber\\
&\qquad=
\lim_{L\rightarrow\infty}
\frac{
\sin[
(\Delta k\Upsilon_\theta+\Delta n\Upsilon_r)L
]
}{
(\Delta k\Upsilon_\theta+\Delta n\Upsilon_r)L
}
=
0,
\end{align}
for
$\Delta k\,\Upsilon_\theta+\Delta n\,\Upsilon_r\neq0$.
Hence only the diagonal contribution survives the long-time average, and
Eq.~\eqref{eq:dKdlambda3b} reduces to
\begin{equation}
\left\langle
\frac{dK^\infty}{d\lambda}
\right\rangle
=
\Gamma
\sum_{\ell mkn}
\frac{
|\check Z^\infty_{\ell mkn}|^2
}{4\pi\omega_{mkn}}
\left(
{\cal M}_{mkn}
+
k\Upsilon_\theta
\right)
\label{eq:dKdlambda4}
\end{equation}
Finally, using the definition of Mino time we have
\begin{equation}
\left\langle
\frac{dK}{d\lambda}
\right\rangle
=
\Gamma
\left\langle
\frac{dK}{dt}
\right\rangle
\end{equation}
together with
\begin{equation}
\frac{dK}{dt}
=
\frac{dQ}{dt}
+
2(aE-L_z)
\left(
a\frac{dE}{dt}
-
\frac{dL_z}{dt}
\right),
\label{eq:Kdot2}
\end{equation}
the Carter flux at infinity is obtained as
\begin{align}
\left\langle
\frac{dQ^\infty}{dt}
\right\rangle \equiv \dot{Q}^{\infty}_{v}
=&
\sum_{\ell mkn}
\frac{\left|Z^{v,\infty}_{\ell mkn}\right|^2}{
4\pi\omega_{mkn}}
\Big[
mL_z\langle\cot^2\theta\rangle
\nonumber\\&-a^2\omega_{mkn}E\langle\cos^2\theta\rangle
+k\Upsilon_\theta
\Big]
\nonumber\\
={}&
2\sum_{\ell mkn}
\frac{
\dot E^{v,\infty}_{\ell mkn}
}{
\omega_{mkn}
}
\left(
{\cal L}_{mkn}+k\Upsilon_\theta
\right),
\label{eq:Qdotnonres_Inf}
\end{align}
An analogous calculation for the ``down'' sector yields the Carter flux at horizon,
\begingroup\small
\begin{align}
\left\langle
\frac{dQ^H}{dt}
\right\rangle
\equiv \dot{Q}^{H}_{v}
&=
\sum_{\ell mkn}
\frac{
\left|Z^{v,H}_{\ell mkn}\right|^2
}{
8\pi Mr_+
\left(\omega_{mkn}-m\Omega_H\right)
}
\nonumber\\
&\times
\Big[
mL_z\langle\cot^2\theta\rangle
-a^2\omega_{mkn}E\langle\cos^2\theta\rangle
+k\Upsilon_\theta
\Big]
\nonumber\\
={}&
2\sum_{\ell mkn}
\frac{
\dot E^{v,H}_{\ell mkn}
}{
\omega_{mkn}
}
\left(
{\cal L}_{mkn}+k\Upsilon_\theta
\right) ,
\label{eq:Qdotnonres_H}
\end{align}
\endgroup
where the factor $Z^{H}_{\ell mkn}$ is defined in Eq.~\eqref{eq:Zcoef} of
Sec.~\ref{sec:fluxes}.  

\section{Adiabatic evolution of inspiral orbit and data analysis}\label{sec:evolution}
\subsection{Adiabatic inspiral model for the inclined and eccentric geodesics}
To compute the full fluxes on generic orbits in Kerr spacetime, besides the electromagnetic fluxes emitted from EMRI system in Eqs.~\eqref{eq:EMtotalflux}, \eqref{eq:Qdotnonres_Inf} and \eqref{eq:Qdotnonres_H}, we also compute gravitational fluxes from EMRI system with the available \texttt{Mathematica}, \texttt{C++} packages in the Black Hole Perturbation Toolkit (\texttt{BHPT})~\cite{BHPToolkit}. In this work, we employ a \texttt{Mathematica}-based package to compute the gravitational energy and angular-momentum fluxes, $(\dot{E},\dot{L},\dot{Q})^{H,\infty}_{g}$, at the Kerr horizon and infinity with arbitrary numerical precision~\cite{BHPTTeukolsky,Hughes:2021exa,Nasipak:2022xjh,Nasipak:2025tby,Zhong:2026jnp}. Once the gravitational and electromagnetic fluxes have been computed, we can obtain the loss rate of orbital integrals using a balance law:
\begin{equation}    
\dot{\cal C}^{\rm orbit}= -\dot{\cal C}_g^{H}- \dot{\cal C}_g^{\infty} -\dot{\cal C}_v^{H}- \dot{\cal C}_v^{\infty}\;,
\end{equation}
where ${\cal C}\in[E,L_z,Q]$. In the adiabatic approximation, the generic inspiral is modeled as a sequence of instantaneous inclined and eccentric geodesics of the Kerr background.
Each geodesic is specified by the semi-latus rectum $p$, eccentricity $e$ and orbital inclination parameter $x$. The dissipative radiation-reaction force induces a slow secular evolution of
these parameters on a timescale much longer than the orbital timescale,
causing $p$, $e$ and $x$ to vary smoothly as energy, angular momentum and Carter constant are carried away by radiation~\cite{Wardell:2021fyy,Hughes:2021exa,German:2023bye}.  Next, we need to find a transformation how to relate rates of change for $p$, $e$ and $x$ to the change rates of $E$, $L_z$ and $Q$, the detailed discussion is given in Appendix B of Ref.~\cite{Hughes:2021exa}.

For a bounded, generic geodesic orbit in Kerr background, there exist two turning points in its radial motion at apoapsis, $r_{\rm a} = p/(1 - e)$, and at periapsis, $r_{\rm peri} = p/(1 + e)$ and it also has a turning point in its polar motion at $\theta_m$.  The radial turning points mean that the radial potential function satisfies $R(r_{\rm a}) = 0$ and $R(r_{\rm peri}) = 0$. The polar turning point means that $\Theta(\theta_m) = 0$. We assume that these conditions hold as the system evolves due to both the backreactions of gravitational and electromagnetic radiation, which means that we require
\begin{eqnarray}\label{eq:Rra:Rrp:Theta:zm}
\frac{d}{dt}R(r_{\rm a}) = 0\;,\quad\frac{d}{dt}R(r_{\rm p}) = 0\;,\quad\frac{d}{dt}\Theta(\theta_m) = 0\;.
\end{eqnarray}
From three equations~\eqref{eq:Rra:Rrp:Theta:zm}, by expanding these total time derivatives and solving a matrix equation, finally we can obtain the change rate of inclination, semi-latus rectum and eccentricity 
\begin{widetext}
\begin{eqnarray}
\frac{dx}{dt} &=& \frac{2x(1 - x^2)L_z(dL_z/dt) - x^3[(dQ/dt) + 2(1 - x^2)a^2 E(dE/dt)]}{2[L_z^2 + x^4a^2(1 - E^2)]}\;,\label{eq:dxIdt2} \\
\frac{dp}{dt} &=& \frac{(1 - e)^2}{2}\frac{dr_{\rm a}}{dt} + \frac{(1 + e)^2}{2}\frac{dr_{\rm p}}{dt} \;,\label{eq:dpdt}\\
\frac{de}{dt} &=& \frac{(1 - e^2)}{2p}\left[(1 - e)\frac{dr_{\rm a}}{dt} - (1 + e)\frac{dr_{\rm p}}{dt}\right]\;,\label{eq:xdot}
\end{eqnarray}
\end{widetext}
where
\begin{equation}
\frac{dr_{\rm a, peri}}{dt} = J_{Er_{\rm a,peri}}\frac{dE}{dt} + J_{L_zr_{\rm a,peri}}\frac{dL_z}{dt} + J_{Qr_{\rm a,peri}}\frac{dQ}{dt}\;.\label{eq:dra:drp}
\end{equation}
Once $dr_{\rm a,p}/dt$ are known, it is not difficult to compute the trajectories  with $dp/dt$ and $de/dt$, the quantities in Eq.~\eqref{eq:dra:drp} are given in Ref~\cite{Hughes:2021exa}
\begingroup\small
\begin{eqnarray}
 J_{Er_{\rm a, peri}} &\equiv& \frac{4aM(L_z - aE)r_{\rm a, peri} - 2Er_{\rm a,peri}^2(a^2 + r_{\rm a, peri}^2)}{{\cal D}(r_{\rm a,peri})}\;,
\nonumber\\
J_{L_zr_{\rm a,peri}} &\equiv& \frac{4M(a E - L_z)r_{\rm a,peri} + 2L_zr_{\rm a,peri}^2}{{\cal D}(r_{\rm a,peri})}\;,
\nonumber\\
J_{Qr_{\rm a,peri}} &\equiv& \frac{r_{\rm a,peri}^2 - 2Mr_{\rm a,peri} + a^2}{{\cal D}(r_{\rm a,peri})}\;,
\end{eqnarray}
\endgroup
with 
\begin{eqnarray}
{\cal D}(r) &\equiv& 2M[Q + (L_z - aE)^2]
\nonumber\\ &&- 2r[L_z^2 + Q + a^2(1 -E^2)]
\nonumber\\
& &+ 6Mr^2 - 4r^3(1 - E^2)\;.
\end{eqnarray}
The constants of orbital motion, $(E, L_z, Q)$, can be mapped to the geometrical orbital parameters $(p,e,x)$, with the corresponding transformation relations derived by Schmidt and van de Meent~\cite{Schmidt:2002qk,vandeMeent:2019cam}.  Consequently, the secular
evolution of the orbital parameters is determined by the rates of change $(dE/dt,dL_z/dt,dQ/dt)$, which are driven by gravitational and electromagnetic radiation reaction.  In the adiabatic approximation, the computational efficiency of a fully relativistic inspiral evolution is therefore governed primarily by the accurate evaluation of the gravitational and electromagnetic fluxes along generic Kerr geodesics. The resulting inspiral can then be represented as a continuous trajectory through the $(p,e,x)$ parameter space.

\subsection{Four-dimensional Chebyshev interpolation}\label{sec:chebyshev}
Modeling an adiabatic inspiral requires the dissipative fluxes to be evaluated repeatedly over a multidimensional domain of generic Kerr geodesics. Direct numerical evaluation of the gravitational and additional-field fluxes at every point encountered during an inspiral is computationally expensive.  We therefore construct surrogate flux
functions using tensor-product Chebyshev interpolation, following the
general strategy of Ref.~\cite{Lynch:2021ogr}.  Chebyshev interpolation is
particularly well suited to this purpose because its nonuniform sampling
reduces the spurious boundary oscillations associated with Runge's
phenomenon while providing rapid and accurate evaluation of smooth
functions throughout the interpolation domain.

The Chebyshev polynomial of the first kind, $T_n(z)$, is defined by $   T_n(\cos\vartheta)=\cos(n\vartheta)$. In the present implementation, we employ Chebyshev--Gauss--Lobatto (CGL) points.  For a grid containing $N$ nodes, the corresponding points on the interval $[-1,1]$ are
\begin{equation}
    z_j=\cos\left(\frac{j\pi}{N-1}\right),
    \qquad
    j=0,\ldots,N-1 .
    \label{eq:CGLnodes}
\end{equation}
Unlike the roots of $T_N$, the CGL grid includes both endpoints of the
interpolation interval.

For generic Kerr orbits, each flux quantity may be regarded as a function
of four orbital variables.  Rather than using the semi-latus rectum $p$
directly, we introduce the frequency-based radial coordinate
\begin{equation}
    u=(1+e)
    \left[
    \frac{\Omega_\phi(a,p,e,x)}
    {\Omega_\phi\left(
       a,
       p_{\rm sep}(a,e,x)/(1+e),
       e,x
    \right)}
    \right]^{2/3},
    \label{eq:u_coordinate}
\end{equation}
where $p_{\rm sep}(a,e,x)$ denotes the separatrix associated with the
corresponding generic Kerr geodesic.  This transformation is useful because
it provides a regular radial coordinate whose upper boundary can be chosen
close to the separatrix while retaining adequate resolution in the
weak-field region.

We therefore construct the interpolation in the four-dimensional parameter
space $(a,u,e,x)$ with $a\in[0,0.9999], u\in[0.08,0.97], e\in[0,0.5], x\in[0.001,1].$
Next each physical coordinate is mapped linearly into the Chebyshev interval $[-1,1]$.  Defining the following parameters
\begin{align}
    \tilde a &=
    \frac{2a-(a_{\max}+a_{\min})}
         {a_{\max}-a_{\min}},
    &
    \tilde u &=
    \frac{2u-(u_{\max}+u_{\min})}
         {u_{\max}-u_{\min}},
    \nonumber\\
    \tilde e &=
    \frac{2e-(e_{\max}+e_{\min})}
         {e_{\max}-e_{\min}},
    &
    \tilde x &=
    \frac{2x-(x_{\max}+x_{\min})}
         {x_{\max}-x_{\min}}\;,
    \label{eq:4D_mapping}
\end{align}
EMRI flux on generic orbit can be approximately expressed with four-dimensional Chebyshev polynomials
$\mathcal{F}$
\begin{align}
    \mathcal{F}(a,u,e,x)
    \simeq
    &\sum_{i=0}^{N_a-1}
    \sum_{j=0}^{N_u-1}
    \sum_{k=0}^{N_e-1}
    \sum_{l=0}^{N_x-1}
    c_{ijkl}\,
   \nonumber \\ &\times  T_i(\tilde a)
    T_j(\tilde u)
    T_k(\tilde e)
    T_l(\tilde x),
    \label{eq:4DChebyshev}
\end{align}
where $c_{ijkl}$ are the four-dimensional Chebyshev coefficients.

In following subsection, we present the implementation details of 4-dimensional interpolation. For sampling points, we take grid point number as
\begin{equation}
    N_a=13,\quad
    N_u=17,\quad
    N_e=13,\quad
    N_x=9\;.
\end{equation}
CGL nodes in the $a$, $u$, $e$ and $x$ directions, respectively.  The
resulting tensor-product grid therefore contains $N_{\rm grid} = N_a N_u N_e N_x  =13\times17\times13\times9 = 25857$ sampling points. In particular, the radial variable $u$ is sampled at 17 CGL nodes over the interval $u\in[0.08,0.97]$.

For each grid point $(a_i,u_j,e_k,x_l)$, the defining relation
\eqref{eq:u_coordinate} is inverted to determine the corresponding
semi-latus rectum, $ p=p(a_i,u_j,e_k,x_l)$.
The gravitational and vector fluxes are then evaluated on the
generic geodesic orbit specified by $(a_i,p,e_k,x_l)$.  In this way, each flux
quantity is tabulated as
\begin{equation}
\mathcal{F}_{ijkl}  =  \mathcal{F}(a_i,u_j,e_k,x_l).
\end{equation}
The coefficients $c_{ijkl}$ in Eq.~\eqref{eq:4DChebyshev} are obtained from
the values $\mathcal{F}_{ijkl}$ using the discrete Chebyshev polynomials.  Numerically, this can be implemented efficiently by successively applying one-dimensional Chebyshev transforms along the four coordinate directions, which is equivalent to a four-dimensional tensor-product discrete cosine transform.  This procedure avoids solving a dense system of $25857$ simultaneous equations and substantially reduces the cost of computing EMRI flux using perturbation theory.
At each point of the sampling grid, the EMRI fluxes from the perturbative theory are computed by summing over the multipolar harmonic indices.
The multipolar sums begin at $\ell_{\min}=1$ for vector perturbations
and at $\ell_{\min}=2$ for gravitational perturbations, while
$\ell_{\max}=10$ is adopted for both sectors. The truncation orders
$n_{\max}$ and $k_{\max}$ are determined independently using an adaptive
convergence criterion. Specifically, each cutoff is increased until three consecutive $n$- or $k$-harmonics change the accumulated flux by a fractional amount smaller than the mass-ratio. 
The convergence of the multipolar sum is further verified by requiring
$\left|\dot{\mathcal F}^{(\ell_{\max})}-\dot{\mathcal F}^{(\ell_{\max}-1)}
\right| \leq10^{-4},$
where $\dot{\mathcal F}^{(\ell_{\max})}$ denotes the total flux accumulated
through multipole order $\ell_{\max}$. These convergence method are applied
separately to the energy, angular-momentum and Carter fluxes at
infinity and at the event horizon for the gravitational and vector fluxes.

Once the coefficients have been determined, evaluating the flux at an arbitrary point $(a,p,e,x)$ requires only three steps: first, the orbital frequency is used to compute $u$ from Eq.~\eqref{eq:u_coordinate}; second, $(a,u,e,x)$ is mapped to $(\tilde a,\tilde u,\tilde e,\tilde x)$ using Eq.~\eqref{eq:4D_mapping}; and finally the finite sum in Eq.~\eqref{eq:4DChebyshev} is evaluated.  The resulting interpolation functions, $(\dot{E},\dot{L},\dot{Q})^{H,\infty,{\rm int}}_{v,g}$, provide the gravitational and vector energy, angular-momentum, and Carter fluxes at a computational cost that is negligible compared with their direct evaluation from the perturbation equations.  These interpolated fluxes can therefore be used efficiently in the adiabatic evolution equations to generate fully relativistic trajectories through the three-dimensional orbital parameter space $(p,e,x)$ for different values of MBH-spin.

To evaluate the validity of this interpolation method, we compare EMRI fluxes, $\dot{\mathcal{C}}_{s,v}^{\infty,H, \rm int}$ and $\dot{\mathcal{C}}_{s,v}^{\infty,H, \rm pert}$ over our sampling grids, the superscript ``int'' denoted by the interpolation functions  and  the superscript ``pert'' denoted by the direct perturbative calculations. We then evaluate the interpolation error by comparing the interpolated and directly computed EMRI fluxes at different points within the orbital parameter space of sampling grids. Additionally, we can assess the precision of vector Carter flux interpolations for any locations in the parameter space of 25857 grid points. Table~\ref{tab:flux_interp} presents four interpolation errors for computing different EMRI fluxes for two sets of orbital parameter configurations $\mathcal{P}_{a,b}$ with different orbital inclination. These errors are also consistent with interpolation error analysis in previous paper~\cite{Speri:2024qak}, which also supports the adiabatic inspiral of orbital parameters. With these fluxes obtained with the Chebyshev-interpolation method, one can generate the gravitational and vector fluxes fast with $\texttt{FEW}$ package, then evolve orbital parameters with Eq.~\eqref{eq:dxIdt2}, \eqref{eq:dpdt} and \eqref{eq:xdot} to obtain the fully relativistic trajectory.

\begin{table*}[htbp]
\centering
\caption{Chebyshev interpolation of the energy and Carter fluxes, and their absolute interpolation errors are listed. Two kinds of EMRI fluxes are computed with the interpolation functions and perturbation theory at any off-grid point, assuming two sets of parameters $\mathcal{P}_a=\{a/M=0.8546,p/M=9.8467, e=0.4328\}$ and $\mathcal{P}_b=\{a/M=0.95,p/M=10.1930405906075, e=0.4081632653061225\}$ keeping same configuration with Ref.~\cite{Speri:2024qak}.}
\begin{tabular}{c|ccc}
\hline\hline
${}$ & Orbital inclination  & Interpolated function & Abs. error
\\\hline
$\mathcal{P}_a$ &1.0   & $\dot{E}_{g}^{H,\rm int}$      &$7.71\times 10^{-4}$ \\
${}$ &1.0     &$\dot{E}_{g}^{\infty,\rm int}$           & $5.84 \times 10^{-4}$ \\
${}$ &1.0    &$\dot{E}_{v}^{H, \rm int}$             & $6.92 \times 10^{-4}$ \\
${}$ &1.0    &$\dot{E}_{v}^{\infty, \rm int}$             & $5.34 \times 10^{-4}$ \\
${}$ &$0.6$  &$\dot{E}_v^{H,\rm int}$        & $5.68 \times 10^{-4}$ \\
${}$ &$0.6$  &$\dot{E}_v^{\infty,\rm int}$     & $7.42 \times 10^{-4}$ \\
${}$ &$0.6$  &$\dot{Q}_v^{H,\rm int}$         & $6.91 \times 10^{-4}$ \\
${}$ &$0.6$  &$\dot{Q}_v^{\infty,\rm int}$     & $5.83 \times 10^{-4}$ \\
${}$ &$0.3$  &$\dot{E}_v^{H,\rm int}$        & $4.54 \times 10^{-4}$ \\
${}$ &$0.3$  &$\dot{E}_v^{\infty,\rm int}$     & $5.67 \times 10^{-4}$ \\
${}$ &$0.3$  &$\dot{Q}_v^{H,\rm int}$         & $6.36 \times 10^{-4}$ \\
${}$ &$0.3$  &$\dot{Q}_v^{\infty,\rm int}$     & $4.29 \times 10^{-4}$ \\
\hline\hline
$\mathcal{P}_b$ & 1.0  &$\dot{E}_{g}^{\infty,\rm int}$  & $8.15 \times 10^{-4}$ \\
${}$  & 1.0 &   $\dot{E}_{g}^{H,\rm int}$                & $7.62 \times 10^{-4}$ \\
${}$ & 1.0  &$\dot{E}_{v}^{H,\rm int}$     & $3.54 \times 10^{-4}$ \\
${}$ & 1.0     &$\dot{E}_{v}^{\infty,\rm int}$                & $2.53 \times 10^{-4}$ \\
${}$ &$0.6$  &$\dot{E}_v^{H,\rm int}$        & $4.76 \times 10^{-4}$ \\
${}$ &$0.6$  &$\dot{E}_v^{\infty,\rm int}$     & $6.51 \times 10^{-4}$ \\
${}$ &$0.6$  &$\dot{Q}_v^{H,\rm int}$         & $7.26 \times 10^{-4}$ \\
${}$ &$0.6$  &$\dot{Q}_v^{\infty,\rm int}$     & $4.73 \times 10^{-4}$ \\
${}$ &$0.3$  &$\dot{E}_v^{H,\rm int}$        & $5.06 \times 10^{-4}$ \\
${}$ &$0.3$  &$\dot{E}_v^{\infty,\rm int}$     & $6.18 \times 10^{-4}$ \\
${}$ &$0.3$  &$\dot{Q}_v^{H,\rm int}$         & $5.12 \times 10^{-4}$ \\
${}$ &$0.3$  &$\dot{Q}_v^{\infty,\rm int}$     & $4.75\times 10^{-4}$ \\
\hline\hline
    \bottomrule
  \end{tabular}
  \label{tab:flux_interp}
\end{table*}

\begin{table}[htbp]
\caption{Prior distributions adopted in the MCMC analyses presented in this work. Barred quantities denote the injected parameter values  to generate the signal shown in MCMC analysis. Throughout the analysis, we fix \(\delta=0.01\). The notation \(U(x_1,x_2)\) denotes a uniform distribution over the interval \([x_1,x_2]\)}.\label{tab:para:injected}
\label{tab:priors}
\centering
\begin{tabular}{cccc}
\toprule\hline\hline
Parameter & Injected value & \qquad & Prior distribution \\	\hline
\midrule
$\ln M$ & $10.49$ & \qquad & $U[(1 - \delta) \cdot \ln \bar{M}, \, (1 + \delta) \cdot \ln \bar{M}]$ \\
$\ln m_p$ & $1.28$ & \qquad & $U[(1 - \delta) \cdot \ln \bar{m}_p, \, (1 + \delta) \cdot \ln  \bar{m}_p]$ \\
$a/M$ & $0.95$ & \qquad & $U[(1 - \delta) \cdot \bar{a}, \, (1 + \delta) \cdot \bar{a}]$ \\
$p/M$ & $30.98$ & \qquad & $U[(1 - \delta) \cdot \bar{p}, \, (1 + \delta) \cdot \bar{p}]$ \\
$e$ & $0.40$ & \qquad & $U[(1 - \delta) \cdot \bar{e}, \, (1 + \delta) \cdot \bar{e}]$ \\
$x$ & $0.50$ & \qquad & $U[0.001, 1.0]$ \\
$D_L[\rm Gpc]$ & $0.37$ & \qquad & $U[0.01, \, 20]$ \\
$\cos \theta_S$ & $0$ & \qquad & $U[-0.99999, \, 0.99999]$ \\
$\phi_S$ & $\pi$ & \qquad & $U[0, \, 2 \pi]$ \\
$\cos \theta_K$ & $1/\sqrt{2}$ & \qquad & $U[-0.99999, \, 0.99999]$ \\
$\phi_K$ & $\pi/3$ & \qquad & $U[0, \, 2 \pi]$ \\
$\Phi_{\phi,0}$ & $\pi/3$ & \qquad & $U[0, \, 2 \pi]$ \\
$\Phi_{r,0}$ & $\pi$ & \qquad & $U[0, \, 2 \pi]$ \\
$\Phi_{\theta,0}$ & $\pi$ & \qquad & $U[0, \, 2 \pi]$ \\
$q_v$ & $0.20$ & \qquad & $U[-1, \, 1]$ \\\hline\hline
\bottomrule
\end{tabular}
\end{table}

\subsection{Waveform model and parameter estimation}\label{wave:constraint}
We compute the tensor and vector fluxes generated by eccentric and inclined Kerr orbits and use them to evolve the corresponding EMRI trajectories in the adiabatic approximation. The observable GW signal is subsequently constructed using the augmented analytical kludge (AAK) waveform model. This hybrid approach combines relativistic orbital evolution with an efficient approximate prescription for waveform generation. In particular, it captures the cumulative dephasing induced by the vector charge through its modification of the orbital energy,  angular-momentum and Carter constant losses. Although a fully relativistic waveform model would improve the quantitative accuracy of the inferred constraints, it is not expected to alter our main qualitative conclusions regarding the detectability of vector-field effects~\cite{Katz:2021yft,Mitra:2025tag}.

The evolution of the orbital elements is obtained by numerically integrating Eqs.~\eqref{eq:dxIdt2}, \eqref{eq:dpdt} and \eqref{eq:xdot}. The resulting trajectories are then supplied to the AAK waveform model implemented in the FastEMRIWaveform (\texttt{FEW}) package. In this construction, the inspiral dynamics are governed by the relativistic tensor and vector fluxes derived above, whereas the emitted tensor waveform is evaluated using the leading-order quadrupole prescription. Consequently, the vector field affects the observed signal primarily through the accumulated orbital phase. The resulting waveform should therefore be regarded as an efficient approximate model rather than a fully relativistic EMRI template. With GPU acceleration, a one-year waveform can be generated on a timescale of order one second, making repeated waveform evaluations in a Bayesian analysis computationally feasible.

In the AAK model, two GW polarizations in the Solar System barycentric frame are expressed as sums over orbital harmonics:
\begingroup\footnotesize
\begin{align}
h_{+} \equiv& \sum_n A_n^{+}
=\sum_n
\Bigg\{
-\left[1+\left(\hat{\boldsymbol L}\cdot\hat{\boldsymbol n}\right)^2\right]
\left[a_n\cos(2\gamma)-b_n\sin(2\gamma)\right]
\nonumber  \\ & +c_n\left[1-\left(\hat{\boldsymbol L}\cdot\hat{\boldsymbol n}\right)^2\right]
\Bigg\},
\nonumber\\
h_{\times}
&\equiv \sum_n A_n^{\times} =\sum_n
2\left(\hat{\boldsymbol L}\cdot\hat{\boldsymbol n}\right)
\left[b_n\cos(2\gamma)+a_n\sin(2\gamma)\right].
\label{amplitude}
\end{align}
\endgroup
Here, \(\hat{\boldsymbol n}\) is the unit vector pointing from the detector toward the source, and \(\hat{\boldsymbol L}\) is the unit vector along the orbital angular momentum. Their relative orientation is specified by the source-location angles \((\theta_S,\phi_S)\) and the orbital-orientation angles \((\theta_L,\phi_L)\). The harmonic coefficients \(a_n\), \(b_n\), and \(c_n\) depend on the orbital eccentricity and are expressed in terms of Bessel functions~\cite{Peters:1963ux}. For the configuration considered here, the angle \(\gamma\), which specifies the direction of pericenter, is related to the accumulated radial phase \(\Psi_r\) through
\begin{equation}
\cos\gamma
=
\cos\left(\frac{\pi}{4}\right)\cos\Psi_r.
\end{equation}
Further details of the waveform convention and the harmonic coefficients are given in Ref.~\cite{Barack:2003fp}.

To model the LISA response beyond the low-frequency approximation, we project the two gravitational-wave polarizations onto the \(A\) and \(E\) channels of the second-generation time-delay-interferometry (TDI) response using \texttt{fastlisaresponse}~\cite{Katz:2022yqe}. This treatment incorporates the orbital motion of the detector constellation, its time-dependent geometry, and finite-arm-length effects.

We perform Bayesian parameter estimation using the parallel-tempered ensemble sampler \texttt{Eryn}~\cite{Karnesis:2023ras}. In the analyses presented below, we employ three temperatures and 16 walkers per temperature. Convergence is assessed by requiring the total length of each chain to exceed \(50\hat{\tau}\), where \(\hat{\tau}\) is the estimated mean integrated autocorrelation time~\cite{Goodman:2010dyf}. Given the TDI data streams
\(\boldsymbol{s}=\{s_A,s_E\}\), Bayes' theorem yields
\begin{equation}
\mathcal{P}(\boldsymbol{\theta}\mid\boldsymbol{s})
=
\frac{
\mathcal{P}(\boldsymbol{s}\mid\boldsymbol{\theta})
\mathcal{P}(\boldsymbol{\theta})
}{
\mathcal{P}(\boldsymbol{s})
}
\propto
\mathcal{P}(\boldsymbol{s}\mid\boldsymbol{\theta})
\mathcal{P}(\boldsymbol{\theta}),
\label{eq:bayes}
\end{equation}
where \(\mathcal{P}(\boldsymbol{\theta})\) is the prior distribution, \(\mathcal{P}(\boldsymbol{s}\mid\boldsymbol{\theta})\) is the likelihood, and $\mathcal{P}(\boldsymbol{s})$ is the Bayesian evidence. The sampled parameter vector is
\begingroup\small
\begin{equation}
\boldsymbol{\theta}
=
\left\{
M,m_p,a,p,e,x,q_v,
\Phi_{\phi,0},\Phi_{r,0},\Phi_{\theta,0},
\theta_S,\phi_S,\theta_L,\phi_L,D_L
\right\}.
\label{eq:parameters:vector}
\end{equation}
\endgroup
Here, the parameter \(q_v\) denotes the vector charge associated with the class of vector--tensor theories considered in this work. The quantities \(\Phi_{\phi,0}\), \(\Phi_{r,0}\) and \(\Phi_{\theta,0}\) are the initial phases of the azimuthal, radial and polar motions, respectively. The angles \((\theta_S,\phi_S)\) specify the sky location of the source, while \((\theta_L,\phi_L)\) determine the orientation of the orbital angular momentum. Finally, $D_L$ denotes the luminosity distance.

Assuming stationary Gaussian noise that is uncorrelated between the \(A\) and \(E\) TDI channels, the log-likelihood is
\begin{equation}
\ln p(\boldsymbol{s}\mid\boldsymbol{\theta})
=
-\frac{1}{2}
\left(
\boldsymbol{s}-\boldsymbol{h}(\boldsymbol{\theta})
\middle|
\boldsymbol{s}-\boldsymbol{h}(\boldsymbol{\theta})
\right)
+\mathrm{const.},
\label{eq:mcmc_likelihood}
\end{equation}
where
\(\boldsymbol{h}(\boldsymbol{\theta})=\{h_A(\boldsymbol{\theta}),
h_E(\boldsymbol{\theta})\}\)
is the waveform projected onto the two TDI channels. The corresponding noise-weighted inner product is defined as
\begin{equation}
\left(\boldsymbol{h}_1\middle|\boldsymbol{h}_2\right)
=
4\,\mathrm{Re}
\sum_{I\in\{A,E\}}
\int_{f_{\rm low}}^{f_{\rm high}}
\frac{
\tilde{h}_{1,I}^{\,*}(f)\tilde{h}_{2,I}(f)
}{
S_{n,I}(f)
}\,df,
\label{eq:inner_product}
\end{equation}
where \(S_{n,I}(f)\) is the one-sided noise power spectral density of TDI channel \(I\), and \(\tilde{h}_{i,I}(f)\) denotes the Fourier transform of \(h_{i,I}(t)\). We adopt the LISA sensitivity model of Ref.~\cite{Robson:2018ifk} and set
\begin{equation}
f_{\rm low}=10^{-4}\,\mathrm{Hz},
\qquad
f_{\rm high}=f_{\rm LSO},
\end{equation}
where \(f_{\rm LSO}\) is the maximum relevant gravitational-wave frequency at the last stable orbit of the Kerr spacetime.

The optimal network signal-to-noise ratio is then
\begin{equation}
\rho_{\rm SNR}
=
\sqrt{
\left(\boldsymbol{h}\middle|\boldsymbol{h}\right)
}.
\end{equation}
This Bayesian analysis yields posterior constraints on the vector charge \(q_v\), together with the intrinsic and extrinsic parameters of the source. The injected parameter values and their corresponding prior distributions are summarized in Table~\ref{tab:priors}.

\begin{table*}[t]
\centering   
\caption{
Carter fluxes at infinity and the event horizon for vector and gravitational perturbations. The fluxes are reported as $\bm{\dot Q^{\infty}}(\dot Q^{H})$, with the asymptotic infinity
results highlighted in boldface and the fluxes at the horizon are enclosed in parentheses. The two column blocks correspond to $(s,\ell,m)=(0,1,1)$ and $(-2,2,2)$,
respectively. For each multipole, the contributions are summed
over the radial and polar harmonics within $-12\leq n\leq12$ and $-12\leq k\leq12$, using 512 orbital points.} \label{tab:vector-gravitational-carter-fluxes}
  \scriptsize
  \setlength{\tabcolsep}{5.2pt}
  \renewcommand{\arraystretch}{1.24}
  \resizebox{\textwidth}{!}{%
  \begin{tabular}{cc|ccc|ccc}
    \toprule\hline\hline
    \multicolumn{2}{c|}{}
      & \multicolumn{3}{c|}{Vector Carter flux, $(s,\ell,m)=(0,1,1)$}
      & \multicolumn{3}{c}{Gravitational Carter flux, $(s,\ell,m)=(-2,2,2)$} \\
    \cmidrule(lr){3-5}\cmidrule(lr){6-8}
    $p/M$ & $\theta_{\rm inc}$ & $e=0.2$ & $e=0.4$ & $e=0.8$ & $e=0.2$ & $e=0.4$ & $e=0.8$ \\\hline
    \midrule
    \multicolumn{8}{c}{$a/M=0.1$} \\\hline
    \midrule
    \multirow{3}{*}{$8$} & $5^\circ$ & \shortstack[c]{$\displaystyle \bm{1.46\times10^{-5}}$\\$\displaystyle \bigl(1.43\times10^{-7}\bigr)$} & \shortstack[c]{$\displaystyle \bm{1.31\times10^{-5}}$\\$\displaystyle \bigl(2.46\times10^{-7}\bigr)$} & \shortstack[c]{$\displaystyle \bm{5.17\times10^{-6}}$\\$\displaystyle \bigl(1.92\times10^{-7}\bigr)$} & \shortstack[c]{$\displaystyle \bm{4.88\times10^{-5}}$\\$\displaystyle \bigl(2.29\times10^{-8}\bigr)$} & \shortstack[c]{$\displaystyle \bm{5.12\times10^{-5}}$\\$\displaystyle \bigl(6.57\times10^{-8}\bigr)$} & \shortstack[c]{$\displaystyle \bm{3.45\times10^{-6}}$\\$\displaystyle \bigl(2.38\times10^{-8}\bigr)$} \\
     & $40^\circ$ & \shortstack[c]{$\displaystyle \bm{7.98\times10^{-4}}$\\$\displaystyle \bigl(9.95\times10^{-6}\bigr)$} & \shortstack[c]{$\displaystyle \bm{7.17\times10^{-4}}$\\$\displaystyle \bigl(1.60\times10^{-5}\bigr)$} & \shortstack[c]{$\displaystyle \bm{2.88\times10^{-4}}$\\$\displaystyle \bigl(1.24\times10^{-5}\bigr)$} & \shortstack[c]{$\displaystyle \bm{1.89\times10^{-3}}$\\$\displaystyle \bigl(1.22\times10^{-6}\bigr)$} & \shortstack[c]{$\displaystyle \bm{2.00\times10^{-3}}$\\$\displaystyle \bigl(3.38\times10^{-6}\bigr)$} & \shortstack[c]{$\displaystyle \bm{1.47\times10^{-4}}$\\$\displaystyle \bigl(8.09\times10^{-7}\bigr)$} \\
     & $85^\circ$ & \shortstack[c]{$\displaystyle \bm{1.96\times10^{-3}}$\\$\displaystyle \bigl(3.97\times10^{-5}\bigr)$} & \shortstack[c]{$\displaystyle \bm{1.78\times10^{-3}}$\\$\displaystyle \bigl(5.81\times10^{-5}\bigr)$} & \shortstack[c]{$\displaystyle \bm{7.52\times10^{-4}}$\\$\displaystyle \bigl(4.69\times10^{-5}\bigr)$} & \shortstack[c]{$\displaystyle \bm{1.79\times10^{-3}}$\\$\displaystyle \bigl(3.03\times10^{-6}\bigr)$} & \shortstack[c]{$\displaystyle \bm{1.93\times10^{-3}}$\\$\displaystyle \bigl(8.63\times10^{-6}\bigr)$} & \shortstack[c]{$\displaystyle \bm{2.16\times10^{-4}}$\\$\displaystyle \bigl(8.86\times10^{-6}\bigr)$} \\
    \midrule
    \multirow{3}{*}{$25$} & $5^\circ$ & \shortstack[c]{$\displaystyle \bm{1.72\times10^{-6}}$\\$\displaystyle \bigl(1.16\times10^{-9}\bigr)$} & \shortstack[c]{$\displaystyle \bm{1.43\times10^{-6}}$\\$\displaystyle \bigl(9.02\times10^{-10}\bigr)$} & \shortstack[c]{$\displaystyle \bm{4.04\times10^{-7}}$\\$\displaystyle \bigl(2.22\times10^{-10}\bigr)$} & \shortstack[c]{$\displaystyle \bm{1.43\times10^{-6}}$\\$\displaystyle \bigl(7.70\times10^{-12}\bigr)$} & \shortstack[c]{$\displaystyle \bm{1.31\times10^{-6}}$\\$\displaystyle \bigl(5.93\times10^{-12}\bigr)$} & \shortstack[c]{$\displaystyle \bm{1.92\times10^{-7}}$\\$\displaystyle \bigl(1.66\times10^{-12}\bigr)$} \\
     & $40^\circ$ & \shortstack[c]{$\displaystyle \bm{9.36\times10^{-5}}$\\$\displaystyle \bigl(4.05\times10^{-8}\bigr)$} & \shortstack[c]{$\displaystyle \bm{7.80\times10^{-5}}$\\$\displaystyle \bigl(2.83\times10^{-8}\bigr)$} & \shortstack[c]{$\displaystyle \bm{2.20\times10^{-5}}$\\$\displaystyle \bigl(6.33\times10^{-9}\bigr)$} & \shortstack[c]{$\displaystyle \bm{5.41\times10^{-5}}$\\$\displaystyle \bigl(1.48\times10^{-10}\bigr)$} & \shortstack[c]{$\displaystyle \bm{4.98\times10^{-5}}$\\$\displaystyle \bigl(4.97\times10^{-11}\bigr)$} & \shortstack[c]{$\displaystyle \bm{7.28\times10^{-6}}$\\$\displaystyle \bigl(1.85\times10^{-10}\bigr)$} \\
     & $85^\circ$ & \shortstack[c]{$\displaystyle \bm{2.25\times10^{-4}}$\\$\displaystyle \bigl(6.54\times10^{-8}\bigr)$} & \shortstack[c]{$\displaystyle \bm{1.88\times10^{-4}}$\\$\displaystyle \bigl(7.98\times10^{-8}\bigr)$} & \shortstack[c]{$\displaystyle \bm{5.32\times10^{-5}}$\\$\displaystyle \bigl(2.57\times10^{-8}\bigr)$} & \shortstack[c]{$\displaystyle \bm{4.86\times10^{-5}}$\\$\displaystyle \bigl(2.44\times10^{-10}\bigr)$} & \shortstack[c]{$\displaystyle \bm{4.49\times10^{-5}}$\\$\displaystyle \bigl(4.40\times10^{-10}\bigr)$} & \shortstack[c]{$\displaystyle \bm{6.60\times10^{-6}}$\\$\displaystyle \bigl(1.68\times10^{-10}\bigr)$} \\\hline
    \midrule
    \multicolumn{8}{c}{$a/M=0.5$} \\\hline
    \midrule
    \multirow{3}{*}{$8$} & $5^\circ$ & \shortstack[c]{$\displaystyle \bm{1.30\times10^{-5}}$\\$\displaystyle \bigl(3.58\times10^{-7}\bigr)$} & \shortstack[c]{$\displaystyle \bm{1.12\times10^{-5}}$\\$\displaystyle \bigl(3.05\times10^{-7}\bigr)$} & \shortstack[c]{$\displaystyle \bm{3.87\times10^{-6}}$\\$\displaystyle \bigl(9.63\times10^{-8}\bigr)$} & \shortstack[c]{$\displaystyle \bm{4.01\times10^{-5}}$\\$\displaystyle \bigl(1.02\times10^{-8}\bigr)$} & \shortstack[c]{$\displaystyle \bm{3.93\times10^{-5}}$\\$\displaystyle \bigl(7.93\times10^{-9}\bigr)$} & \shortstack[c]{$\displaystyle \bm{4.00\times10^{-6}}$\\$\displaystyle \bigl(6.04\times10^{-8}\bigr)$} \\
     & $40^\circ$ & \shortstack[c]{$\displaystyle \bm{7.25\times10^{-4}}$\\$\displaystyle \bigl(-1.01\times10^{-5}\bigr)$} & \shortstack[c]{$\displaystyle \bm{6.32\times10^{-4}}$\\$\displaystyle \bigl(6.42\times10^{-6}\bigr)$} & \shortstack[c]{$\displaystyle \bm{2.23\times10^{-4}}$\\$\displaystyle \bigl(1.33\times10^{-7}\bigr)$} & \shortstack[c]{$\displaystyle \bm{1.66\times10^{-3}}$\\$\displaystyle \bigl(7.57\times10^{-7}\bigr)$} & \shortstack[c]{$\displaystyle \bm{1.66\times10^{-3}}$\\$\displaystyle \bigl(2.28\times10^{-6}\bigr)$} & \shortstack[c]{$\displaystyle \bm{1.40\times10^{-4}}$\\$\displaystyle \bigl(6.03\times10^{-6}\bigr)$} \\
     & $85^\circ$ & \shortstack[c]{$\displaystyle \bm{1.92\times10^{-3}}$\\$\displaystyle \bigl(4.82\times10^{-5}\bigr)$} & \shortstack[c]{$\displaystyle \bm{1.72\times10^{-3}}$\\$\displaystyle \bigl(6.79\times10^{-5}\bigr)$} & \shortstack[c]{$\displaystyle \bm{6.73\times10^{-4}}$\\$\displaystyle \bigl(5.44\times10^{-5}\bigr)$} & \shortstack[c]{$\displaystyle \bm{1.80\times10^{-3}}$\\$\displaystyle \bigl(6.86\times10^{-6}\bigr)$} & \shortstack[c]{$\displaystyle \bm{1.92\times10^{-3}}$\\$\displaystyle \bigl(1.70\times10^{-5}\bigr)$} & \shortstack[c]{$\displaystyle \bm{1.90\times10^{-4}}$\\$\displaystyle \bigl(2.53\times10^{-5}\bigr)$} \\
    \midrule
    \multirow{3}{*}{$25$} & $5^\circ$ & \shortstack[c]{$\displaystyle \bm{1.68\times10^{-6}}$\\$\displaystyle \bigl(8.02\times10^{-9}\bigr)$} & \shortstack[c]{$\displaystyle \bm{1.40\times10^{-6}}$\\$\displaystyle \bigl(7.03\times10^{-9}\bigr)$} & \shortstack[c]{$\displaystyle \bm{6.77\times10^{-7}}$\\$\displaystyle \bigl(1.89\times10^{-9}\bigr)$} & \shortstack[c]{$\displaystyle \bm{1.39\times10^{-6}}$\\$\displaystyle \bigl(9.43\times10^{-11}\bigr)$} & \shortstack[c]{$\displaystyle \bm{1.27\times10^{-6}}$\\$\displaystyle \bigl(9.55\times10^{-11}\bigr)$} & \shortstack[c]{$\displaystyle \bm{1.90\times10^{-7}}$\\$\displaystyle \bigl(9.16\times10^{-12}\bigr)$} \\
     & $40^\circ$ & \shortstack[c]{$\displaystyle \bm{9.21\times10^{-5}}$\\$\displaystyle \bigl(3.24\times10^{-7}\bigr)$} & \shortstack[c]{$\displaystyle \bm{7.66\times10^{-5}}$\\$\displaystyle \bigl(2.81\times10^{-7}\bigr)$} & \shortstack[c]{$\displaystyle \bm{2.14\times10^{-5}}$\\$\displaystyle \bigl(7.45\times10^{-8}\bigr)$} & \shortstack[c]{$\displaystyle \bm{5.32\times10^{-5}}$\\$\displaystyle \bigl(2.46\times10^{-9}\bigr)$} & \shortstack[c]{$\displaystyle \bm{4.89\times10^{-5}}$\\$\displaystyle \bigl(2.33\times10^{-9}\bigr)$} & \shortstack[c]{$\displaystyle \bm{7.21\times10^{-6}}$\\$\displaystyle \bigl(1.28\times10^{-9}\bigr)$} \\
     & $85^\circ$ & \shortstack[c]{$\displaystyle \bm{2.25\times10^{-4}}$\\$\displaystyle \bigl(2.69\times10^{-8}\bigr)$} & \shortstack[c]{$\displaystyle \bm{1.88\times10^{-4}}$\\$\displaystyle \bigl(5.51\times10^{-8}\bigr)$} & \shortstack[c]{$\displaystyle \bm{5.29\times10^{-5}}$\\$\displaystyle \bigl(2.33\times10^{-8}\bigr)$} & \shortstack[c]{$\displaystyle \bm{4.88\times10^{-5}}$\\$\displaystyle \bigl(4.88\times10^{-10}\bigr)$} & \shortstack[c]{$\displaystyle \bm{4.50\times10^{-5}}$\\$\displaystyle \bigl(1.04\times10^{-9}\bigr)$} & \shortstack[c]{$\displaystyle \bm{6.59\times10^{-6}}$\\$\displaystyle \bigl(9.13\times10^{-10}\bigr)$} \\\hline
    \midrule
    \multicolumn{8}{c}{$a/M=0.9$} \\ \hline
    \midrule
    \multirow{3}{*}{$8$} & $5^\circ$ & \shortstack[c]{$\displaystyle \bm{1.17\times10^{-5}}$\\$\displaystyle \bigl(6.48\times10^{-7}\bigr)$} & \shortstack[c]{$\displaystyle \bm{9.89\times10^{-6}}$\\$\displaystyle \bigl(5.67\times10^{-7}\bigr)$} & \shortstack[c]{$\displaystyle \bm{3.16\times10^{-6}}$\\$\displaystyle \bigl(1.79\times10^{-7}\bigr)$} & \shortstack[c]{$\displaystyle \bm{3.39\times10^{-5}}$\\$\displaystyle \bigl(8.16\times10^{-8}\bigr)$} & \shortstack[c]{$\displaystyle \bm{3.19\times10^{-5}}$\\$\displaystyle \bigl(5.23\times10^{-8}\bigr)$} & \shortstack[c]{$\displaystyle \bm{4.51\times10^{-6}}$\\$\displaystyle \bigl(1.01\times10^{-7}\bigr)$} \\
     & $40^\circ$ & \shortstack[c]{$\displaystyle \bm{6.62\times10^{-4}}$\\$\displaystyle \bigl(2.03\times10^{-5}\bigr)$} & \shortstack[c]{$\displaystyle \bm{5.66\times10^{-4}}$\\$\displaystyle \bigl(1.53\times10^{-5}\bigr)$} & \shortstack[c]{$\displaystyle \bm{1.85\times10^{-4}}$\\$\displaystyle \bigl(2.83\times10^{-6}\bigr)$} & \shortstack[c]{$\displaystyle \bm{1.47\times10^{-3}}$\\$\displaystyle \bigl(3.44\times10^{-7}\bigr)$} & \shortstack[c]{$\displaystyle \bm{1.42\times10^{-3}}$\\$\displaystyle \bigl(2.05\times10^{-6}\bigr)$} & \shortstack[c]{$\displaystyle \bm{1.65\times10^{-4}}$\\$\displaystyle \bigl(1.02\times10^{-5}\bigr)$} \\
     & $85^\circ$ & \shortstack[c]{$\displaystyle \bm{1.86\times10^{-3}}$\\$\displaystyle \bigl(7.19\times10^{-5}\bigr)$} & \shortstack[c]{$\displaystyle \bm{1.65\times10^{-3}}$\\$\displaystyle \bigl(9.49\times10^{-5}\bigr)$} & \shortstack[c]{$\displaystyle \bm{5.93\times10^{-4}}$\\$\displaystyle \bigl(6.90\times10^{-5}\bigr)$} & \shortstack[c]{$\displaystyle \bm{1.80\times10^{-3}}$\\$\displaystyle \bigl(1.10\times10^{-5}\bigr)$} & \shortstack[c]{$\displaystyle \bm{1.90\times10^{-3}}$\\$\displaystyle \bigl(2.59\times10^{-5}\bigr)$} & \shortstack[c]{$\displaystyle \bm{1.86\times10^{-4}}$\\$\displaystyle \bigl(4.37\times10^{-5}\bigr)$} \\
    \midrule
    \multirow{3}{*}{$25$} & $5^\circ$ & \shortstack[c]{$\displaystyle \bm{1.65\times10^{-6}}$\\$\displaystyle \bigl(1.44\times10^{-8}\bigr)$} & \shortstack[c]{$\displaystyle \bm{1.37\times10^{-6}}$\\$\displaystyle \bigl(1.26\times10^{-8}\bigr)$} & \shortstack[c]{$\displaystyle \bm{4.18\times10^{-7}}$\\$\displaystyle \bigl(3.36\times10^{-9}\bigr)$} & \shortstack[c]{$\displaystyle \bm{1.35\times10^{-6}}$\\$\displaystyle \bigl(3.43\times10^{-10}\bigr)$} & \shortstack[c]{$\displaystyle \bm{1.23\times10^{-6}}$\\$\displaystyle \bigl(3.54\times10^{-10}\bigr)$} & \shortstack[c]{$\displaystyle \bm{1.87\times10^{-7}}$\\$\displaystyle \bigl(1.79\times10^{-11}\bigr)$} \\
     & $40^\circ$ & \shortstack[c]{$\displaystyle \bm{9.06\times10^{-5}}$\\$\displaystyle \bigl(5.82\times10^{-7}\bigr)$} & \shortstack[c]{$\displaystyle \bm{7.52\times10^{-5}}$\\$\displaystyle \bigl(-5.05\times10^{-7}\bigr)$} & \shortstack[c]{$\displaystyle \bm{2.09\times10^{-5}}$\\$\displaystyle \bigl(-1.33\times10^{-7}\bigr)$} & \shortstack[c]{$\displaystyle \bm{5.24\times10^{-5}}$\\$\displaystyle \bigl(9.28\times10^{-9}\bigr)$} & \shortstack[c]{$\displaystyle \bm{4.80\times10^{-5}}$\\$\displaystyle \bigl(9.19\times10^{-9}\bigr)$} & \shortstack[c]{$\displaystyle \bm{7.15\times10^{-6}}$\\$\displaystyle \bigl(3.82\times10^{-9}\bigr)$} \\
     & $85^\circ$ & \shortstack[c]{$\displaystyle \bm{2.24\times10^{-4}}$\\$\displaystyle \bigl(3.59\times10^{-8}\bigr)$} & \shortstack[c]{$\displaystyle \bm{1.87\times10^{-4}}$\\$\displaystyle \bigl(8.39\times10^{-8}\bigr)$} & \shortstack[c]{$\displaystyle \bm{2.84\times10^{-5}}$\\$\displaystyle \bigl(4.09\times10^{-8}\bigr)$} & \shortstack[c]{$\displaystyle \bm{4.88\times10^{-5}}$\\$\displaystyle \bigl(2.38\times10^{-10}\bigr)$} & \shortstack[c]{$\displaystyle \bm{4.51\times10^{-5}}$\\$\displaystyle \bigl(1.23\times10^{-9}\bigr)$} & \shortstack[c]{$\displaystyle \bm{6.58\times10^{-6}}$\\$\displaystyle \bigl(2.60\times10^{-9}\bigr)$} \\
    \bottomrule\hline\hline
  \end{tabular}%
  }
\end{table*}

\begin{figure*}[htb!]
\centering
\includegraphics[width=\textwidth]{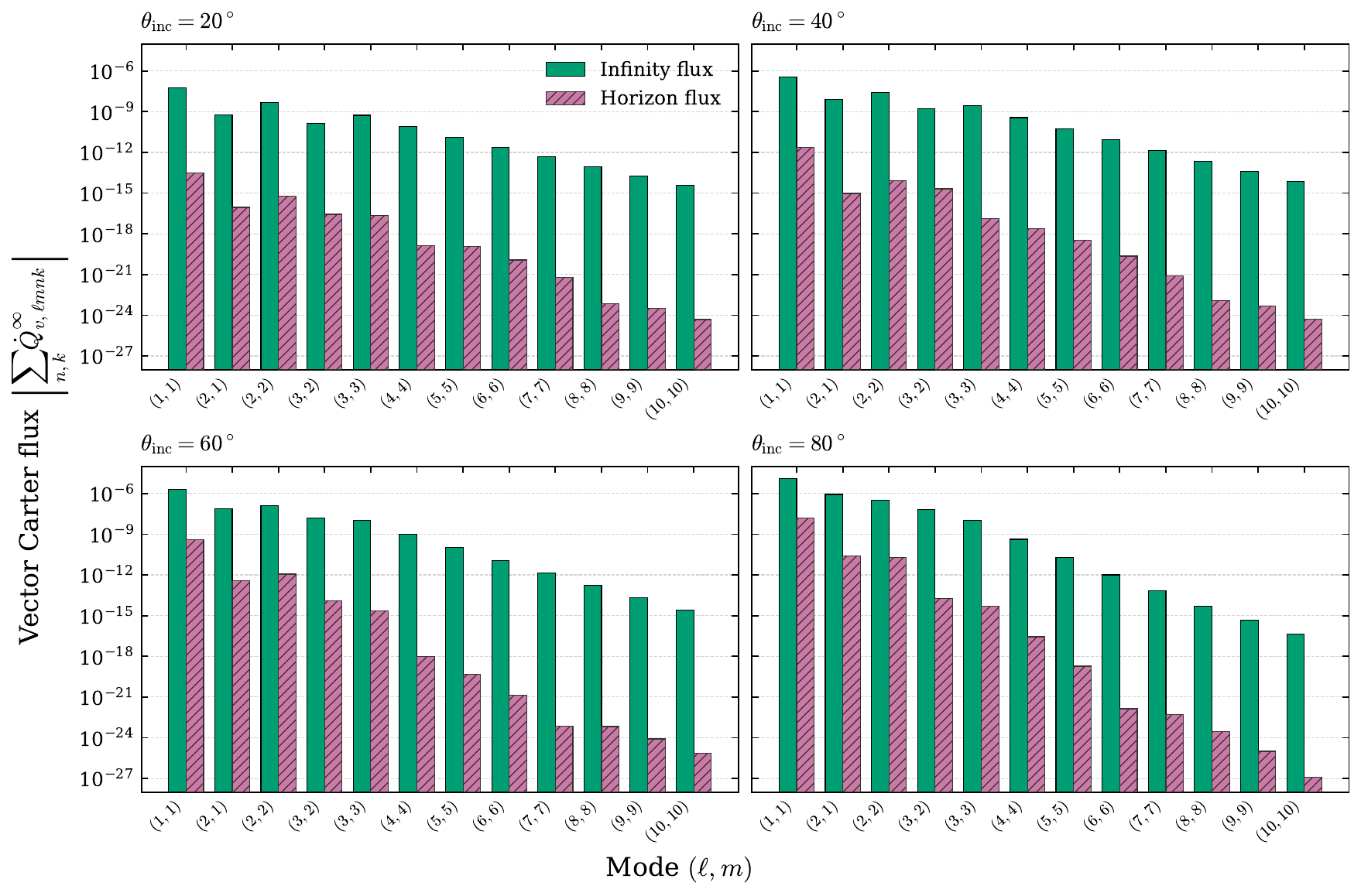}
\caption{Bar plots of the vector Carter fluxes (in units of mass-ratio $\epsilon^2$) at infinity for the vector harmonic modes with some specified angular indices $(\ell, m)$ are outlined, summed over the $n$ harmonics. Other parameters are taken as follows: $a/M=0.9, p/M=6.0, e=0.1, q_v=0.1$.
}\label{fig:bar:carter:flux}
\end{figure*}

\begin{figure}[htb!]
\centering
\includegraphics[width=\linewidth]{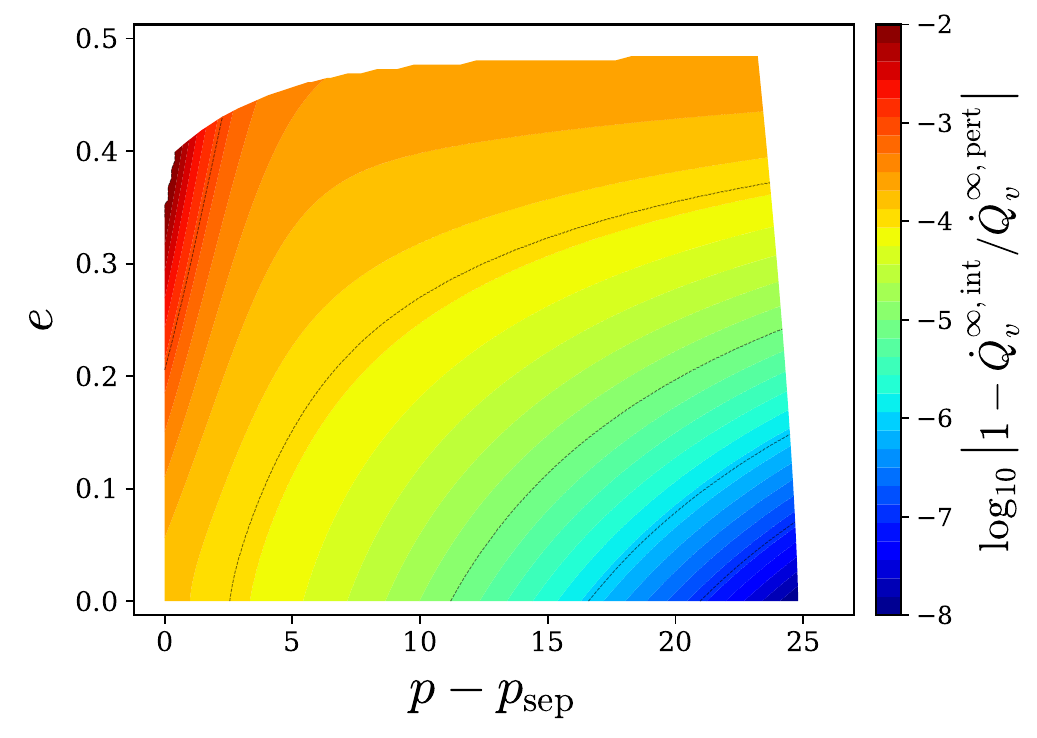}
\caption{Relative error of vector Carter flux from Chebyshev-interpolated method as a contour of parameters $p-p_{\rm sep}$ and $e$ is plotted for vector charge $q_v=0.1$. Other intrinsic parameters are set as follows: $a/M=0.6$, $\theta_{\rm inc}=\pi/5$ and mass-ratio $\epsilon=10^{-5}$.}\label{fig:vectorflux:error}
\end{figure}

\begin{figure*}[htb!]
\centering
\includegraphics[width=\linewidth,trim=1.0 1.0 1.0 1.0,clip]{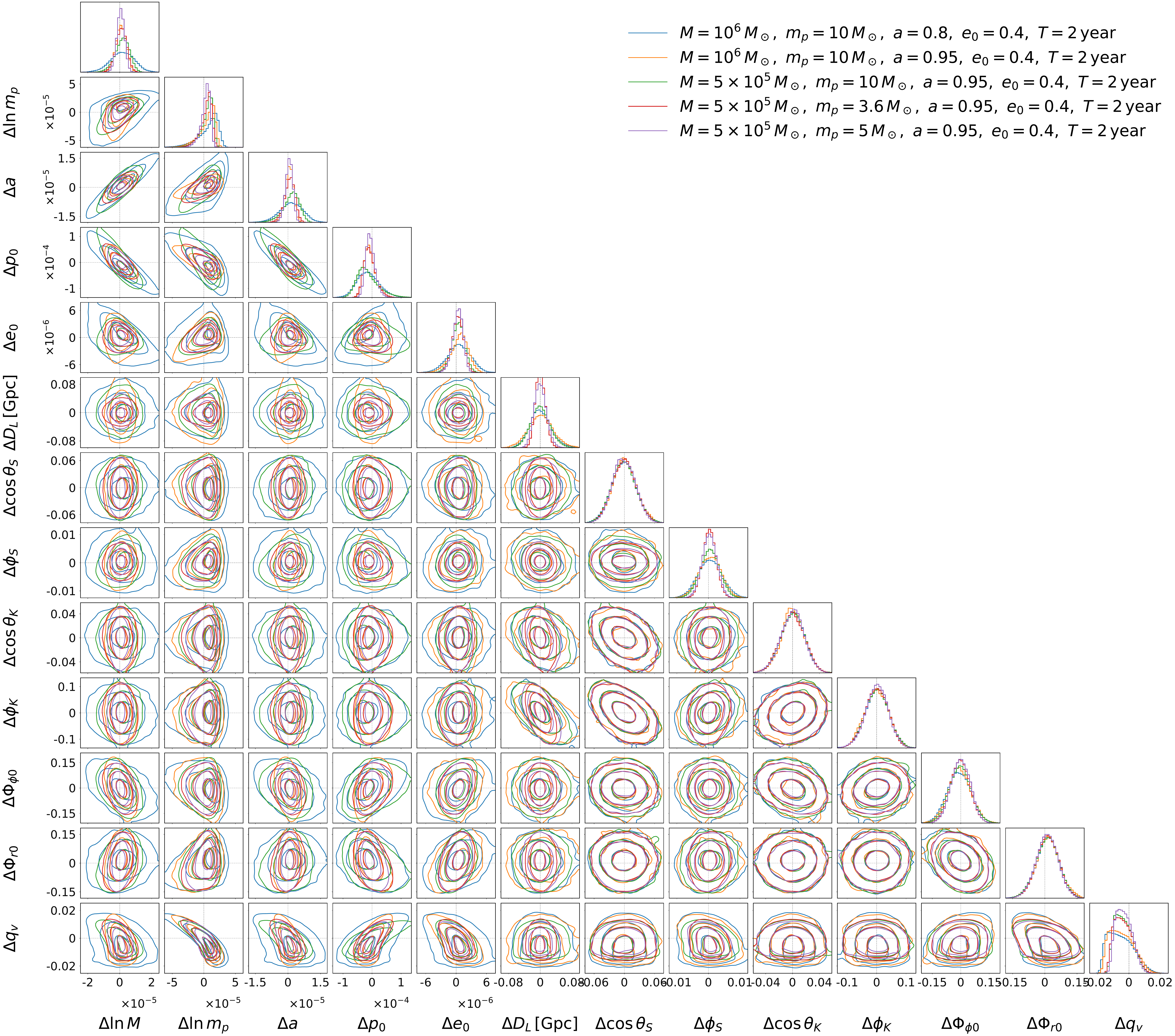}
\caption{Marginalized posterior distributions for five eccentric, equatorial EMRI configurations. All systems have an initial eccentricity of $e_0=0.4$ and an observation duration of
$T=2\,\mathrm{yr}$. The blue and orange contours correspond to $(M/M_\odot,m_p/M_\odot,a/M)=(10^6,10,0.8)$ and $(10^6,10,0.95)$, respectively. The green, red and purple contours correspond to $(5\times10^5,10,0.95)$, $(5\times10^5,3.6,0.95)$ and $(5\times10^5,5,0.95)$,
respectively. The diagonal panels display the one-dimensional marginalized posterior distributions, while the lower-triangular panels show the corresponding two-dimensional joint posterior contours.
}\label{fig:MCMC:equatorial}
\end{figure*}

\begin{figure*}[htb!]
\centering
\includegraphics[width=\linewidth,trim=1.0 1.0 1.0 1.0,clip]{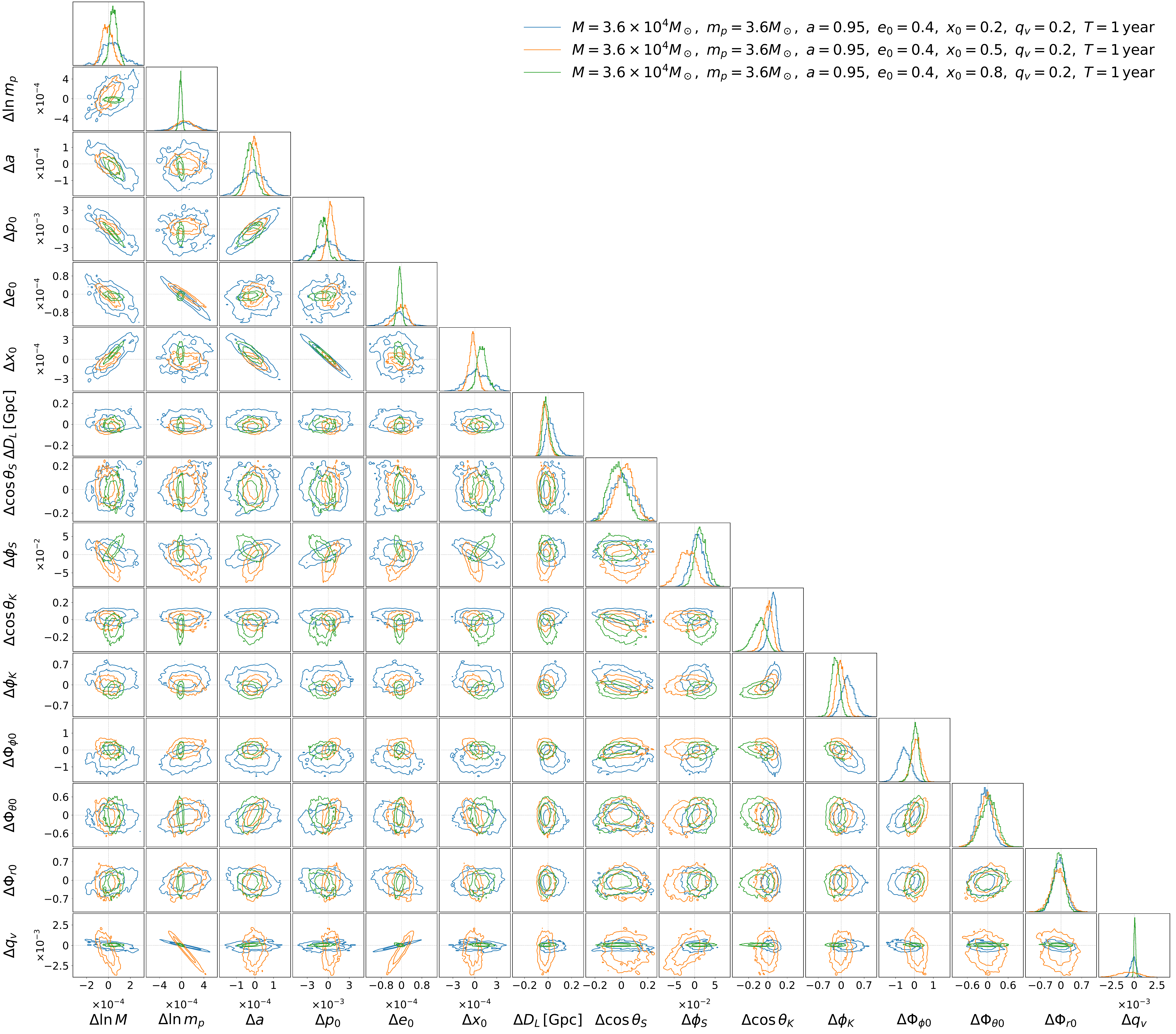}
\caption{Marginalized posterior distributions for a 15-parameter EMRI
model in the vicinity of the injected parameter values. The common source parameters are
$M=3.6\times10^{4}M_\odot$, $m_p=3.6M_\odot$, $a/M=0.95$, $e_0=0.4$ and $q_v=0.2$. All signals have an observation duration of
$T\simeq1\,\mathrm{yr}$, a sampling interval of $\Delta t=10\,\mathrm{s}$, and an injected signal-to-noise
ratio of $\rho=11$. The blue, orange and green contours correspond to
initial inclination parameters $x_0=0.2$, $0.5$ and $0.8$, respectively.
The diagonal panels display the one-dimensional marginalized posterior distributions, while the
lower-triangular panels show the corresponding two-dimensional highest-posterior-density credible
regions containing $68.27\%$ and $95.45\%$ of the posterior probability.
}\label{fig:emri_local_posterior_inclination}
\end{figure*}

\begin{figure*}[htb!]
\centering
\includegraphics[width=\linewidth,trim=1.0 1.0 1.0 1.0,clip]{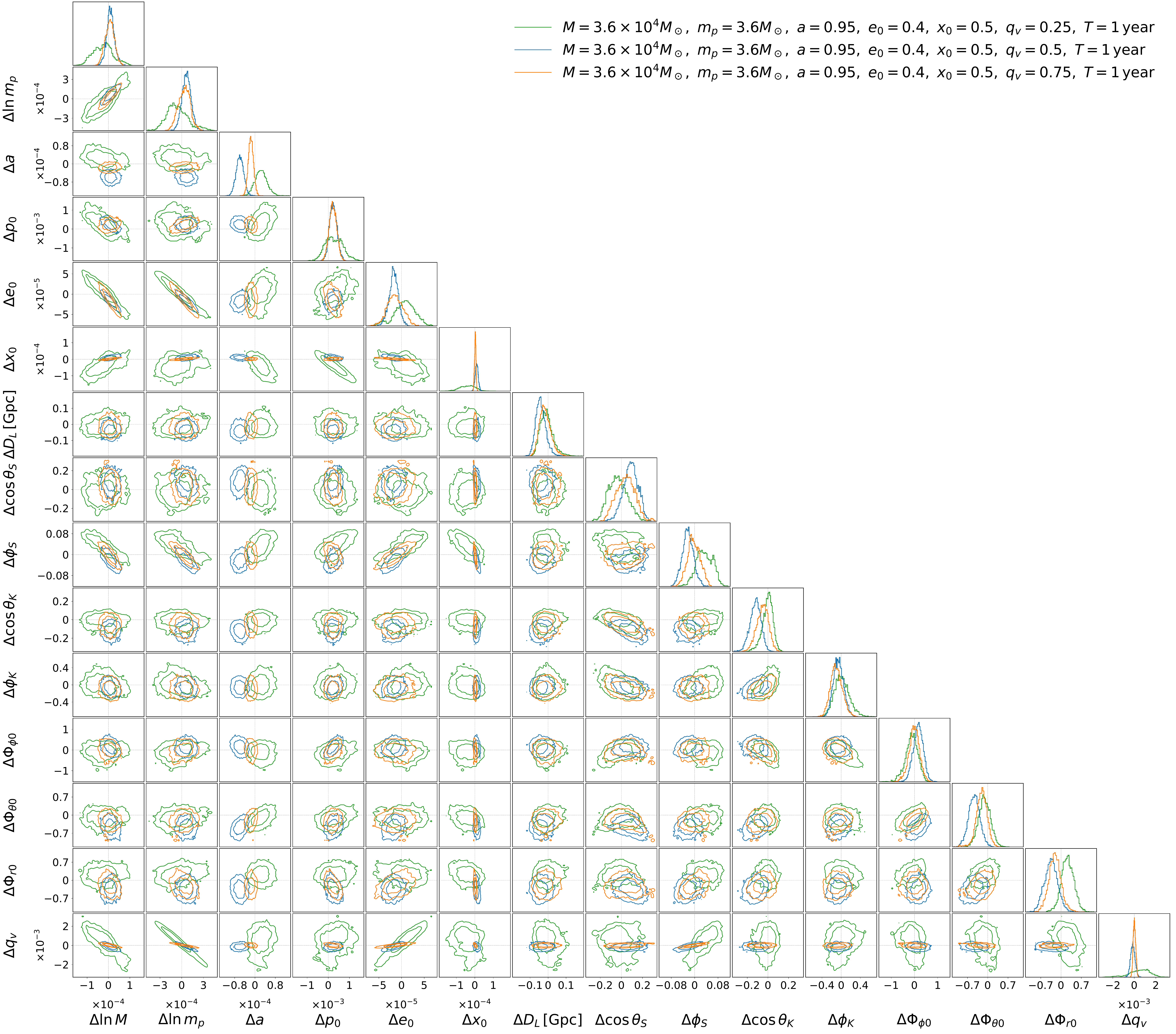}
\caption{Marginalized posterior distributions for a 15-parameter
EMRI model in the vicinity of the injected parameter values.
The common source parameters are
$M=3.6\times10^{4}M_\odot$,
$m_p=3.6M_\odot$, $a/M=0.95$,
$e_0=0.4$ and $x_0=0.5$.
All signals have an observation duration of
$T\simeq1\,\mathrm{yr}$, a sampling interval of
$\Delta t=10\,\mathrm{s}$, and an injected signal-to-noise
ratio of $\rho=11$.
The green, blue and orange contours correspond to
injected vector charges of $q_v=0.25$, $0.5$ and $0.75$,
respectively.
The diagonal panels display the one-dimensional
marginalized posterior distributions, while the
lower-triangular panels show the corresponding
two-dimensional joint posterior contours.
}\label{fig:qv025_050_075_discard500_overlay_reference_style}
\end{figure*}

\begin{figure*}[htb!]
\centering
\includegraphics[width=\linewidth,trim=1.0 1.0 1.0 1.0,clip]{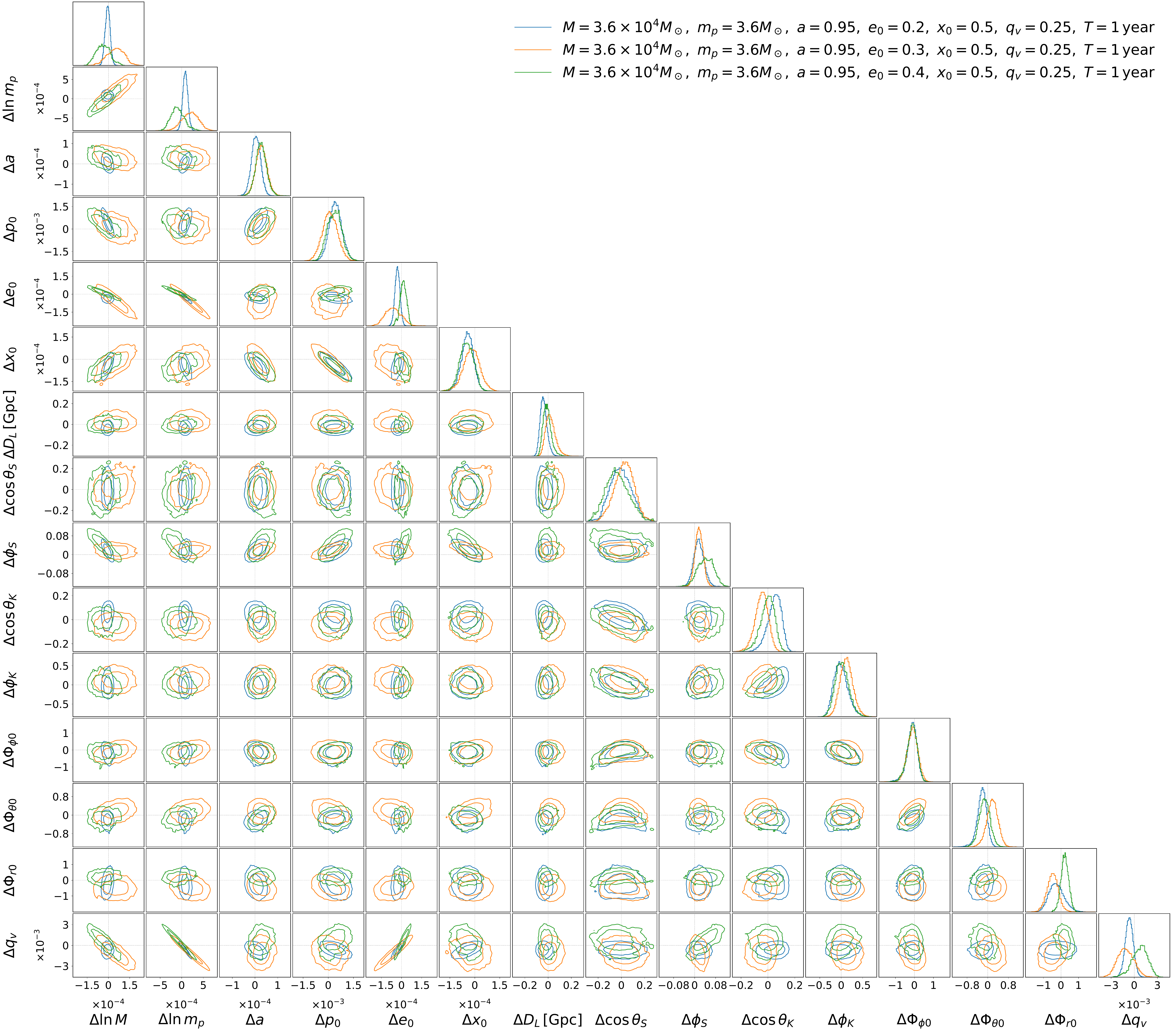}
\caption{Marginalized posterior distributions for a 15-parameter
EMRI model in the vicinity of the injected parameter values.
The common source parameters are $M=3.6\times10^{4}M_\odot$,
$m_p=3.6M_\odot$, $a/M=0.95$ and $x_0=0.5$.
All signals have an observation duration of $T\simeq1\,\mathrm{yr}$, a sampling interval of
$\Delta t=10\,\mathrm{s}$, and an injected signal-to-noise ratio of $\rho=11$. The blue,orange and  green contours correspond to initial eccentricities of $e_0=0.2$, $0.3$ and $0.4$,
respectively. The diagonal panels display the one-dimensional marginalized posterior distributions, while the lower-triangular panels show the corresponding two-dimensional joint posterior contours.
}\label{fig:corner:plot:eccs}
\end{figure*}

\begin{figure*}[htb!]
\centering
\includegraphics[width=.4843\textwidth]{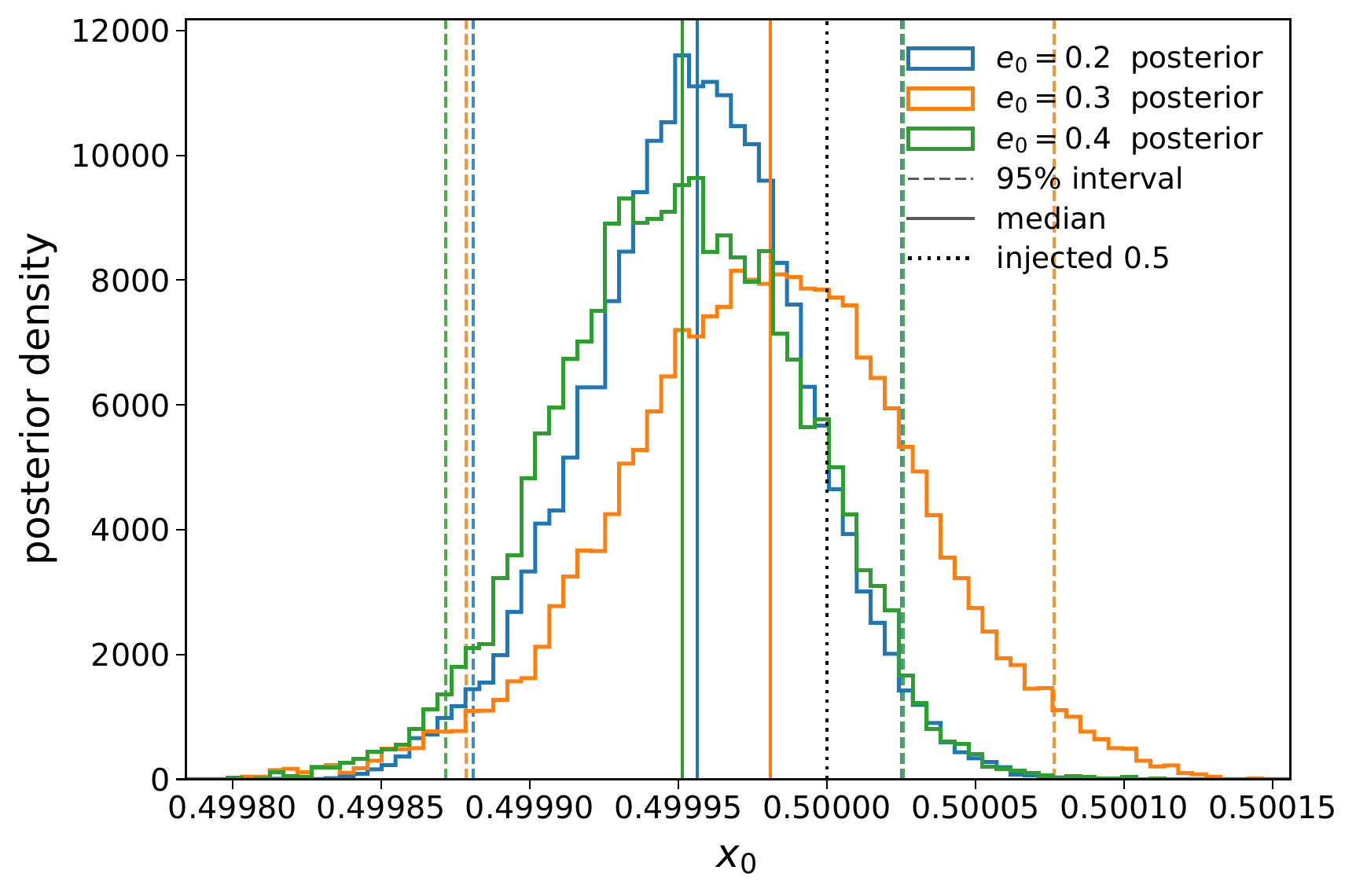}
\includegraphics[width=.4843\textwidth]{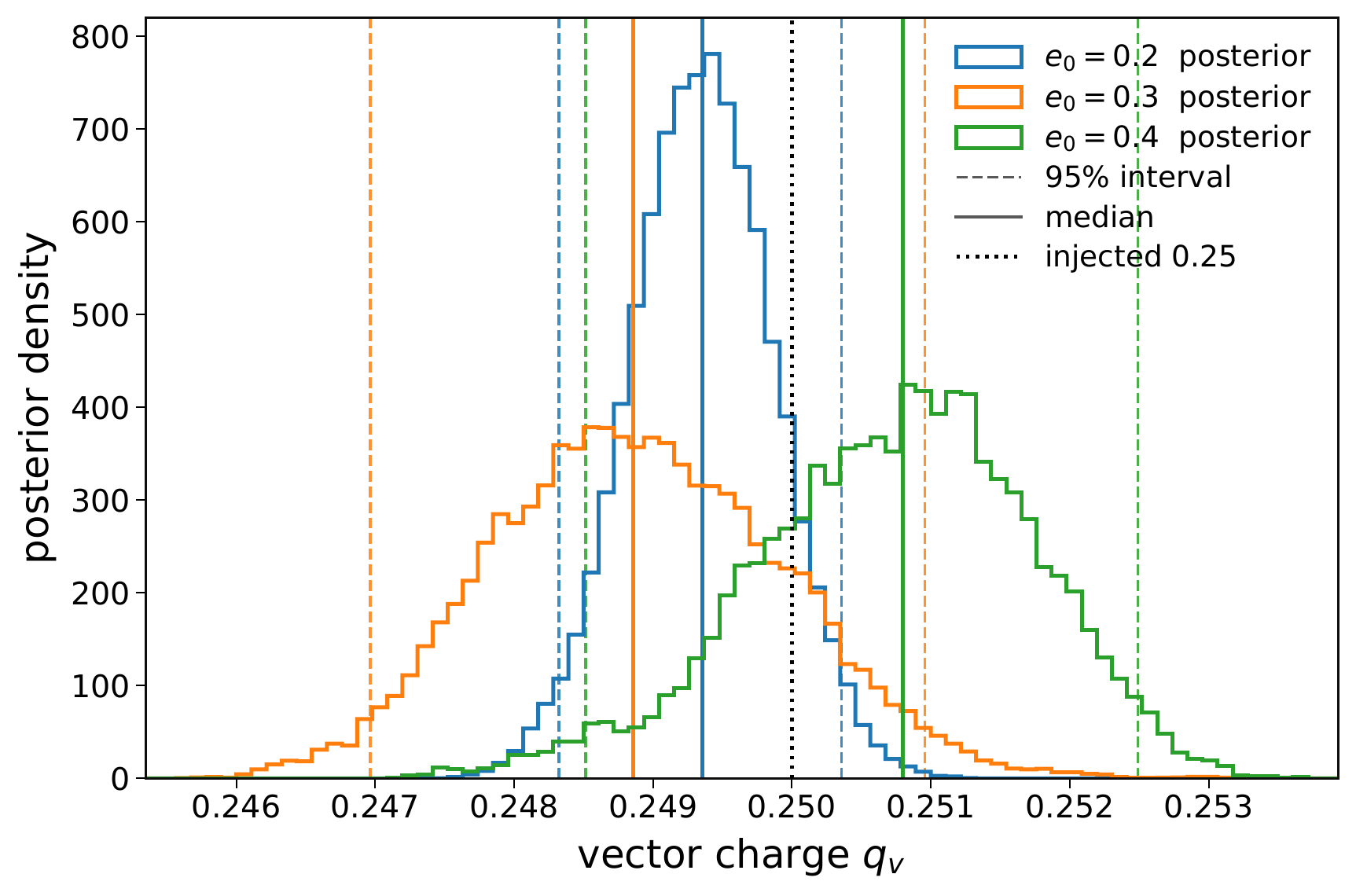}
\caption{The marginalized posterior distributions of orbital inclination parameter $x_0$ (left panel) and the vector charge $q_v$ (right panel) are plotted, with the injected true values $q_v=0.25$ and $x_0=0.5$. The injection source parameters are the same as Fig.~\ref{fig:corner:plot:eccs}.  The estimated median and $95\%$ credible interval are labeled by the dashed lines with different colors.}\label{fig:posterior:distribu:qv:x0}
\end{figure*}

\section{Result}\label{sec:result}
In this section, we present some results of EMRI fluxes on the generic Kerr orbits in subsection~\ref{sec:vector_generic_flux} and the constraint on vector charge using Bayesian analysis in subsection~\ref{sec:vector_mcmc}.

\subsection{Vector fluxes for eccentric and inclined orbits}\label{sec:vector_generic_flux}
In this subsection, we consider the EMRI fluxes on the eccentric and inclined EMRI orbits.  To compute the gravitational fluxes, we employ the Python package \texttt{pybhpt}, which provides a frequency-domain framework for studying perturbations of Kerr spacetime, including metric perturbations and self-force quantities~\cite{Nasipak:2021qfu,Nasipak:2022xjh,Nasipak:2023kuf,
Nasipak:2025tby}.  As discussed in previous studies~\cite{Zhang:2022hbt,Zhang:2023vok,German:2023bye,Torres:2020fye,Zi:2025qos}, when both orbital eccentricity and inclination are simultaneously taken into account, the harmonic structure of the vector fluxes becomes considerably richer. To illustrate the corresponding mode structure, in Fig.~\ref{fig:bar:carter:flux} we present bar plots of the dominant contributions to the vector energy and Carter fluxes, after summing over the harmonic indices $(\ell,m)$.  We find that the overall magnitude of the vector Carter flux associated with different modes decreases gradually as the harmonic indices $(\ell,m)$ increase.  In addition, the contribution of the dominant mode to the vector Carter flux increases mildly with increasing orbital inclination angle $\theta_{\rm inc}$.  These features are useful for the construction of multidimensional flux grids for the evolution of asymmetric binaries, since they provide practical guidance for choosing appropriate truncation and summation strategies for the harmonic modes.

For vector perturbations generated by generic bound geodesics, we develop a new module to evaluate the corresponding fluxes by summing over the full four-index mode decomposition $(\ell,m,n,k)$ for the eccentric and inclined trajectories.  In Table~\ref{tab:vector-gravitational-carter-fluxes}, we report the Carter fluxes at the horizon and at infinity for both vector and gravitational perturbations.  We consider the dominant multipoles $(\ell,m)=(1,1)$ for the vector field and $(\ell,m)=(2,2)$ for the gravitational field, together with the representative orbital configurations
\begin{align}
    a/M &\in \{0.1,0.5,0.9\}, \qquad
    p/M \in \{8,25\},
    \nonumber\\
    e &\in \{0.2,0.4,0.8\}, \qquad
    \theta_{\rm inc}\in\{5^\circ,40^\circ,85^\circ\}.
\end{align}
Since the energy and angular-momentum fluxes exhibit very similar qualitative
behavior~\cite{Zhang:2022hbt,Zhang:2023vok,German:2023bye,Torres:2020fye,Zi:2025qos}, we focus on the rates of change of the energy and Carter constant in Table~\ref{tab:vector-gravitational-carter-fluxes}.

For larger values of the semi-latus rectum $p$, the Carter flux obtained after summing over the $(n,k)$ harmonics decreases, as expected for orbits located farther from the primary black hole.  This behavior is explicitly illustrated in Table~\ref{tab:vector-gravitational-carter-fluxes}, where the  horizon and infinity Carter fluxes,$\dot{Q}_{v,g}^{H,\infty}$, are listed for the orbital configurations under consideration.  For different orbital inclinations, the overall magnitude of the vector Carter flux varies only moderately and remains larger than its gravitational counterpart for the configurations considered here.  We also find that the Carter fluxes increase mildly as either the orbital eccentricity or the spin of the massive black hole is increased.

For the actual evolution of the orbital parameters, repeatedly evaluating the perturbative fluxes at every point along an inspiral would be computationally expensive.  We therefore employ the six interpolation functions $(\dot{E},\dot{L},\dot{Q})^{H,\infty,{\rm int}}_{v,g}$ to evaluate the gravitational and vector fluxes efficiently along generic geodesics and thereby determine the secular evolution of the orbital geometrical parameters.  The interpolation is constructed from the
Chebyshev coefficients $c_{ijkl}$ introduced in Eq.~\eqref{eq:4DChebyshev}, using the EMRI fluxes evaluated on the four-dimensional sampling grid.

To assess the accuracy of this interpolation procedure, in Table~\ref{tab:flux_interp} we compare the interpolated gravitational and vector fluxes with those obtained directly from black-hole perturbation theory at two representative sampling points, ${\cal P}_a$ and ${\cal P}_b$.  The absolute differences between the interpolated and perturbative results are typically of order $10^{-4}$, indicating that the Chebyshev interpolation provides an accurate and efficient representation of the EMRI fluxes at the sampling points.

We further examine the performance of the interpolants away from the sampling nodes in Fig.~\ref{fig:vectorflux:error}.  For this comparison, we fix the black-hole spin to $a=0.6$ and the orbital inclination to $\theta_{\rm inc}=\pi/5$.  The results show that the relative difference
between the interpolated vector Carter flux and the direct perturbative
calculation decreases as the secondary approaches the separatrix and as the
orbital eccentricity increases.  Over most of the orbital parameter region
considered here, the interpolation error remains at approximately the
$10^{-4}$ level.  These results demonstrate that the six Chebyshev-based
interpolation functions $(\dot{E},\dot{L},\dot{Q})^{H,\infty,{\rm int}}_{v,g}$ provide an efficient representation of the gravitational and vector fluxes and can therefore be employed to generate the adiabatic evolution of generic inspiral orbits.

\subsection{MCMC constraint on vector charge}\label{sec:vector_mcmc}
In this subsection, we investigate the constraints on the vector charge using gravitational-wave signals from eccentric and inclined EMRIs within the Bayesian framework defined by Eq.~\eqref{eq:mcmc_likelihood}. We employ the \texttt{Eryn} package to sample the posterior distributions of the source parameters using the MCMC method. The parameter space adopted in the \texttt{FEW} waveform model, together with the corresponding
injected values, is summarized in Table~\ref{tab:para:injected}.
The resulting waveforms are projected onto the LISA detector response
and transformed into the frequency domain for likelihood evaluation.
To suppress spectral leakage associated with the finite observation
duration, we apply a Tukey window with a taper parameter of
$\alpha_{\rm T}=0.05$ before performing the Fourier transform.
The Fourier transforms and likelihood evaluations are implemented
using the GPU-accelerated \texttt{CuPy} library. The details of the
prior distributions and sampling procedure are given below.

We assign a uniform prior to the luminosity distance,
over the interval $D_L\in[0.01,20]\,{\rm Gpc}$. The priors on the intrinsic source
parameters are centered around their respective injected values,
while the initial radial and polar phases have uniform priors over
$\Phi_{r,0},\Phi_{\theta,0}\in[0,2\pi)$.
For all configurations considered in this work, the resulting
marginalized posterior distributions are substantially narrower
than the corresponding prior ranges, indicating that the likelihood
provides appreciable constraints within the adopted parameter
space.

Particular attention is paid to the prior and posterior distributions
of the vector charge $q_v$ carried by the secondary. At leading
order in the adiabatic approximation, the additional vector
energy, angular-momentum and Carter fluxes scale quadratically
with the vector charge. The resulting modification of the orbital
evolution, and hence the accumulated gravitational-wave phase,
therefore depends primarily on $q_v^2$. When the waveform
difference is dominated by the accumulated phase shift, the
charged waveform may be approximated as
\begin{equation}
h(t;q_v)
\simeq
h(t;0)\exp\left[i\Delta\Phi(t;q_v)\right],
\label{eq:waveform_vector_phase}
\end{equation}
where $\Delta\Phi$ represents the phase correction induced by the
additional vector-radiation channel. For sufficiently small phase
differences, this gives
\begin{equation}
h(t;q_v)-h(t;0)
\simeq
i\,h(t;0)\Delta\Phi(t;q_v),
\qquad
\Delta\Phi(t;q_v)\propto q_v^2.
\label{eq:waveform_vector_difference}
\end{equation}
Consequently, the likelihood can exhibit an approximately Gaussian
dependence on $q_v^2$ in the regime where the waveform response is
well described by this leading-order approximation. We adopt a
uniform prior over $[-1,1]$ that includes both positive and negative values of
$q_v$. Since the leading-order dissipative corrections are invariant
under $q_v\rightarrow -q_v$, the corresponding likelihood cannot
distinguish the two charge signs within this approximation.
For each MCMC analysis, we employ 16 walkers and monitor the
integrated autocorrelation time, $\hat{\tau}$, to assess the
convergence of the chains. We require the number of iterations
$N_{\rm it}$ to exceed approximately $50\hat{\tau}$ and verify that
the estimated autocorrelation time has stabilized as the chains
evolve. These criteria are used to assess whether the chains have
explored the posterior distributions sufficiently for the
subsequent parameter-estimation analysis.

Figures~\ref{fig:MCMC:equatorial}--\ref{fig:qv025_050_075_discard500_overlay_reference_style}
show the marginalized posterior distributions of the intrinsic and extrinsic source parameters obtained from our MCMC analyses. The corresponding injected parameter values are given in Table~\ref{tab:para:injected}. These results allow us to examine the dependence of the posterior distributions, particularly that of the vector charge, on the orbital configuration and the intrinsic properties of the EMRI system. We first consider eccentric equatorial inspirals in
Fig.~\ref{fig:MCMC:equatorial}, where five EMRI signals are analyzed with an observation time of two years and an initial eccentricity of $e_0=0.4$. The five configurations differ in the component masses and black-hole spin. The blue and orange contours correspond to
$a/M=0.8$ and $a/M=0.95$, respectively. The green contours represent
a system with $M=5\times10^5 M_\odot$, $m_p=10 M_\odot$ and
$a/M=0.95$. For the red and purple contours, the mass of the primary
is kept fixed at $M=5\times10^5 M_\odot$, while the secondary masses
are chosen to be $m_p=3.6 M_\odot$ and $m_p=5 M_\odot$,
respectively. The joint posterior distributions in the bottom row of Fig.~\ref{fig:MCMC:equatorial} reveal appreciable correlations between the vector charge $q_v$ and the intrinsic source parameters. Specifically, $q_v$ exhibits negative correlations with $\ln m_p$, $e_0$ and $\Phi_{r,0}$, whereas its correlations with the black-hole spin $a$ and initial semi-latus rectum $p_0$ are positive. These correlations indicate that variations in the vector charge can be partially
compensated by changes in other intrinsic parameters when fitting the waveform. Consequently, the accuracy of the vector-charge measurement is closely connected to the joint inference of the component masses, black-hole spin and initial orbital parameters.

We further explore the dependence of parameter estimation on the initial orbital inclination, eccentricity and vector charge. Figure~\ref{fig:emri_local_posterior_inclination} shows the posterior
distributions obtained for three initial inclinations, $x_0\in\{0.2,0.4,0.8\}$, with the remaining parameters fixed at $M=3.6\times10^4 M_\odot$, $m_p=3.6 M_\odot$, $a/M=0.95$,
$e_0=0.4$ and $q_v=0.2$. To investigate the effects of the injected vector charge and initial eccentricity separately, we consider additional configurations with $q_v\in\{0.25,0.5,0.75\}$ and $e_0\in\{0.2,0.3,0.4\}$, respectively. The vector-charge prior is uniform over $[-1,1]$ for all these configurations. Their posterior distributions are displayed in
Figs.~\ref{fig:qv025_050_075_discard500_overlay_reference_style} and~\ref{fig:corner:plot:eccs}. Throughout these analyses, we adopt a one-year observation period and adjust the luminosity
distance $D_L$ to maintain a fixed signal-to-noise ratio of $\rho=11$.

As shown in Fig.~\ref{fig:emri_local_posterior_inclination},
the precision with which $q_v$ is inferred exhibits a nonmonotonic dependence on the initial inclination. Although $q_v$ remains strongly anticorrelated with $\ln m_p$ and
positively correlated with $e_0$, the corresponding joint posterior contours change appreciably as $x_0$ varies. Figure~\ref{fig:qv025_050_075_discard500_overlay_reference_style}
further indicates that increasing the injected vector charge leads to progressively tighter constraints on $q_v$. Likewise, the results in Fig.~\ref{fig:corner:plot:eccs} show that
larger initial eccentricities yield narrower posterior distributions for the source parameters, indicating improved measurement precision for the configurations of more eccentric orbits.

An important feature shared by these three analyses is the persistent correlation structure involving the vector charge. Specifically, $q_v$ exhibits negative correlations with $\ln M$, $\ln m_p$ and $x_0$, while its correlation with $e_0$ remains positive. These trends are maintained across the different initial configurations, even though the widths and shapes of the posterior distributions vary. Taken together, the results indicate that the orbital configuration and injected vector charge influence the precision of Bayesian parameter estimation, while the qualitative correlations between $q_v$ and the intrinsic source parameters remain consistent for all Figs.~\ref{fig:MCMC:equatorial}--\ref{fig:posterior:distribu:qv:x0} considered here.

Finally, in Fig.~\ref{fig:posterior:distribu:qv:x0}, we compare the marginalized posterior distributions of the vector charge $q_v$ and initial inclination parameter $x_0$ for three initial eccentricities, $e_0\in\{0.2,0.3,0.4\}$. The remaining source parameters are fixed to those of the $q_v=0.25$
configuration in Fig.~\ref{fig:qv025_050_075_discard500_overlay_reference_style},
with an injected inclination of $x_0=0.5$. The estimated medians and corresponding $95\%$ credible intervals for the vector charge are $q_v=0.2492_{-0.0009}^{+0.0012}$,
$q_v=0.2488_{-0.0018}^{+0.0022}$ and $q_v=0.2508_{-0.0023}^{+0.0017}$ for $e_0=0.2$, $0.3$ and $0.4$, respectively. For the same eccentricity configurations, the corresponding
estimates of the initial inclination parameter are
$x_0=0.499965_{-0.000084}^{+0.000060}$,
$x_0=0.499956_{-0.000077}^{+0.000121}$ and
$x_0=0.49997_{-0.000098}^{+0.000056}$,
respectively. These results illustrate how the marginalized constraints on $q_v$ and $x_0$ vary with the initial orbital eccentricity for the source parameter configurations considered here.

\section{Conclusion}\label{sec:conclusion}
In this work, we have investigated the influence of an additional massless vector field on extreme-mass-ratio inspirals with eccentric and inclined orbits around a Kerr black hole. We consider a system
in which the massive primary is described by a Kerr spacetime, while the secondary carries an effective vector charge $q_v$. The additional vector radiation modifies the secular evolution of the orbital energy, angular momentum and Carter constant. Our main theoretical contribution is the derivation of the orbit-averaged Carter-constant evolution driven by electromagnetic radiation for generic, nonresonant Kerr
geodesics. We express the corresponding contributions at infinity and the event horizon in terms of frequency-domain electromagnetic modes. Together with the energy and angular-momentum fluxes,
this result provides the evolution rates of three constants to determine the adiabatic evolution of eccentric and inclined geodesic orbits.

We have subsequently combined the electromagnetic contributions with the corresponding gravitational fluxes to determine the evolution of the orbital parameters $(p,e,x)$. The simultaneous radial and polar motions introduce a four-index harmonic structure $(\ell,m,n,k)$, making direct flux evaluation along an inspiral computationally demanding. To address this difficulty, we have constructed four-dimensional Chebyshev interpolants for the gravitational and vector fluxes. Our numerical calculations illustrate the dependence of the Carter-constant evolution on the orbital eccentricity, inclination, semi-latus rectum and black-hole spin. The interpolation framework provides an efficient means of evaluating these dissipative contributions during the generation of adiabatic inspiral trajectories.

To investigate the implications for EMRI observations, we have incorporated the flux-driven orbital evolution into an approximate waveform model and performed Bayesian parameter estimation for representative EMRI orbital configurations. The resulting posterior distributions show that the inference of the vector charge depends on the intrinsic properties of the source and its initial orbital configuration. In particular, the vector charge exhibits appreciable correlations with the component masses and orbital parameters, while the inclination-dependent constraints display a nonmonotonic behavior
for the configurations considered. Variations in the injected vector-charge also modify the widths and correlation structure of the posterior distributions. These results highlight the importance
of accounting for both radial and polar orbital motion when assessing the parameter-estimation implications of an additional vector-radiation channel.

The orbital evolution is restricted to the leading adiabatic order and nonresonant geodesics, while the gravitational-wave signal is generated using an approximate waveform prescription. Furthermore, the leading dissipative vector corrections depend quadratically on $q_v$ and are therefore insensitive to its sign within this framework. A more complete analysis will require assessing waveform systematics, incorporating post-adiabatic effects and orbital resonances, and examining the possible degeneracies between vector radiation and astrophysical environmental effects. Connecting the effective vector charge to the coupling parameters of a specific gravitational theory will also be necessary for translating the present phenomenological constraints into theory-dependent bounds.

The framework developed here connects electromagnetic perturbation theory, the adiabatic evolution of generic Kerr orbits, and Bayesian gravitational-wave inference. It provides a basis frame for investigating how additional radiative degrees of freedom can influence the long-term orbital dynamics and parameter estimation of EMRIs in future space-based gravitational-wave observations. Additionally, several extensions of the present waveform model are required for a more comprehensive assessment of vector-charge constraints. First, the inclusion of post-adiabatic corrections will be important for
accounting for higher-order effects in the mass-ratio expansion and improving the accuracy of the inspiral dynamics~\cite{Spiers:2023cva}. A related extension is to incorporate the spin of the secondary and its coupling to the orbital motion along generic Kerr trajectories~\cite{Burke:2023lno}.

Second, realistic EMRIs may live in astrophysical environments, including accretion disks and dark-matter distributions, whose gravitational and dissipative effects can modify the orbital
evolution and the accumulated GW phase~\cite{Kejriwal:2023djc,Zi:2025onl,Copparoni:2025vty,Zi:2026zpw}.
A systematic treatment of these effects, together with higher-order relativistic corrections and additional
vector radiation, will be necessary to quantify potential parameter degeneracies and distinguish the signatures of vector charge from those of the astrophysical environment. These developments will enable a more robust assessment of vector-charge constraints from future space-based GW observations.

\section{Acknowledgments}
T. Z. also thanks Lorenzo Speri for providing a very detailed explanation on interpolation techniques in Mathematica. The work is funded by the National Natural Science Foundation of China with Grant No.~12405059, the Startup Fund for Advanced Talents of Nanchang University with Grant No. 28170256, Key Program of the Natural Science Foundation of Jiangxi Province under Grant No. 20232ACB201008, the Ganpo Juncai Innovative Talent Program and the Ganpo High-Level Innovative Talent Program,
the Startup Fund for Advanced Talents of Putian University  Grant No. 2023141, Scientific Research (on Science and Technology) Projects for Young and Middle-Aged Teachers in Fujian province  Grant No. JZ240055, and the Natural Science Foundation of Fujian Province  Grant No. 2025J011042.
\clearpage
\bibliographystyle{apsrev4-2}
\bibliography{References}

\end{document}